\documentclass[aps,prd,reprint,longbibliography,showpacs,showkeys]{revtex4-1}
\usepackage{amsmath}
\usepackage{amssymb}
\usepackage{graphicx}
\usepackage[unicode=true,pdfusetitle,
 bookmarks=true,bookmarksnumbered=false,bookmarksopen=false,
 breaklinks=false,pdfborder={0 0 1},backref=false,colorlinks=false]
 {hyperref}
\allowdisplaybreaks[1]

\makeatletter

\newcommand{\noun}[1]{\textsc{#1}}

\@ifundefined{textcolor}{}
{%
 \definecolor{BLACK}{gray}{0}
 \definecolor{WHITE}{gray}{1}
 \definecolor{RED}{rgb}{1,0,0}
 \definecolor{GREEN}{rgb}{0,1,0}
 \definecolor{BLUE}{rgb}{0,0,1}
 \definecolor{CYAN}{cmyk}{1,0,0,0}
 \definecolor{MAGENTA}{cmyk}{0,1,0,0}
 \definecolor{YELLOW}{cmyk}{0,0,1,0}
}
\numberwithin{equation}{section}
\numberwithin{figure}{section}

\makeatother

\begin{document}

\title{Symmetry improvement of 3PI effective actions for $\mathrm{O}\left(N\right)$
scalar field theory}

\author{Michael J. \surname{Brown}}

\affiliation{College of Science, Technology and Engineering, James Cook University,
Townsville 4811, Australia}

\email{michael.brown6@my.jcu.edu.au}

\author{Ian B. \surname{Whittingham}}

\affiliation{College of Science, Technology and Engineering, James Cook University,
Townsville 4811, Australia}

\email{ian.whittingham@jcu.edu.au}

\date{February 12, 2015}
\begin{abstract}
n-Particle Irreducible Effective Actions (nPIEA) are a powerful tool
for extracting non-perturbative and non-equilibrium physics from quantum
field theories. Unfortunately, practical truncations of nPIEA can
unphysically violate symmetries. Pilaftsis and Teresi (PT) addressed
this by introducing a ``symmetry improvement'' scheme in the context
of the 2PIEA for an $\mathrm{O}\left(2\right)$ scalar theory, ensuring
that the Goldstone boson is massless in the broken symmetry phase
{[}A. Pilaftsis and D. Teresi, Nuclear Physics B 874, 2 (2013), pp.
594--619.{]}. We extend this idea by introducing a symmetry improved
3PIEA for $\mathrm{\mathrm{O}}\left(N\right)$ theories, for which
the basic variables are the one-, two- and three-point correlation
functions. This requires the imposition of a Ward identity involving
the three-point function. We find that the method leads to an infinity
of physically distinct schemes, though a field theoretic analogue
of d'Alembert's principle is used to single out a unique scheme. The
standard equivalence hierarchy of nPIEA no longer holds with symmetry
improvement and we investigate the difference between the symmetry
improved 3PIEA and 2PIEA. We present renormalized equations of motion
and counter-terms for two and three loop truncations of the effective
action, though we leave their numerical solution to future work. We
solve the Hartree-Fock approximation and find that our method achieves
a middle ground between the unimproved 2PIEA and PT methods. The phase
transition predicted by our method is weakly first order and the Goldstone
theorem is satisfied, while the PT method correctly predicts a second
order phase transition. In contrast, the unimproved 2PIEA predicts
a strong first order transition with large violations of the Goldstone
theorem. We also show that, in contrast to PT, the two loop truncation
of the symmetry improved 3PIEA does not predict the correct Higgs
decay rate although the three loop truncation does, at least to leading
order. These results suggest that symmetry improvement should not
be applied to $n$PIEA truncated to $<n$ loops. We also show that
symmetry improvement schemes are compatible with the Coleman-Mermin-Wagner
theorem, giving a check on the consistency of the formalism.
\end{abstract}

\maketitle

\section{Introduction\label{sec:Introduction}}

The recent demands of non-equilibrium field theory applications in
particle physics, cosmology and condensed matter have led to a renaissance
in the development of novel field theory methods. The $S$-matrix
school, rebooted in the guise of spinor-helicity methods, has led
to a dramatic speedup in the computation of gauge theory scattering
amplitudes in vacuum \citep{*[{There is a vast literature. For a pedagogical review see e.g. }] [{and for an implementation see e.g. }] Elvang2013,*Berger2008}.
On the finite temperature and density fronts, efficient functional
methods in the form of $n$-particle irreducible effective actions
($n$PIEA) have proven useful to understand collective behaviour and
phase transitions \citep{Berges2004}. They are similar in spirit
to methods based on Schwinger-Dyson equations in field theory or BBGKY
(Bogoliubov-Born-Green-Kirkwood-Yvon) equations in kinetic theory
however, unlike the Schwinger-Dyson or BBGKY equations, $n$PIEA naturally
form closed systems of equations of motion without requiring any closure
ansatz \citep{Carrington2004,Carrington2011a,Carrington2011}. $n$PIEA
methods can be understood as a hybrid of variational and perturbative
methods: $n$PIEA consist of a series of Feynman diagrams, however
the propagators and vertices of these diagrams are the \emph{exact}
1- through $n$-point proper connected correlation functions which
are determined self-consistently using variational equations of motion.

\begin{figure}
\includegraphics[width=1\columnwidth]{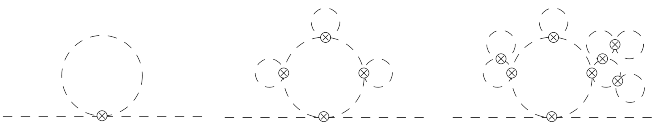}

\protect\caption{\label{fig:hartree-superdaisies}From left to right: the Hartree-Fock
self-energy diagram, an example daisy or ring diagram, an example
super-daisy graph. The whole class of super-daisy diagrams is obtained
from iterating insertions of Hartree-Fock graphs in all possible ways.}
\end{figure}

This self-consistency effectively resums certain classes of perturbative
Feynman diagrams to infinite order. For example, the one loop 2PIEA
diagram corresponding to the Hartree-Fock self-energy in $\phi^{4}$
theory actually sums all of the so-called daisy and super-daisy graphs
of ordinary perturbation theory (Figure \ref{fig:hartree-superdaisies}).
This particular resummation is often done in the literature without
the use of $n$PIEA, but such \emph{ad hoc} resummation schemes run the risk
of summing an asymptotic series: a mathematically dangerous operation
(recent progress on summability has been made in \emph{resurgence}
theory \citep{*[{See e.g. }] [{}] Cherman2014,*Dorigoni2014}, which is beyond
the scope of this work). $n$PIEA sidestep this issue because they
are defined by the rigorous Legendre transform procedure, guaranteeing
equivalence with the original theory. Unlike \emph{ad hoc} resummations,
$n$PIEA based approximation schemes are placed on a firm theoretical
footing and can be systematically improved.

However, loop-wise truncations of $n$PIEA, $n>1$, have difficulties
in the treatment of theories with spontaneously broken continuous
symmetries. The root cause of these difficulties is the fact that
$n$PIEA obey different Ward identities than the $1$PIEA. When the
effective action is truncated to a finite order the equivalence between
the Ward identities is lost. This can also be understood in terms
of the resummation of perturbative Feynman diagrams: when an $n$PIEA
is truncated some subset of perturbative diagrams are summed to infinite
order, but the complementary subset is left out entirely. The pattern
of resummations does not guarantee that the cancellations between
perturbative diagrams needed to maintain the symmetry are kept. In
the case of scalar field theories with $\mathrm{O}\left(N\right)\to\mathrm{O}\left(N-1\right)$
breaking, the result is that the final $\mathrm{O}\left(N-1\right)$
symmetry is maintained, but, at the Hartree-Fock level of approximation,
the nonlinearly realised $\mathrm{O}\left(N\right)/\mathrm{O}\left(N-1\right)$
is lost, the Goldstone theorem is violated (the $N-1$ Goldstone bosons
are massive), and the symmetry restoration phase transition is first
order in contradiction with the second order transition expected on
the basis of universality arguments. A similar problem arises in
gauge theories, where the violation of gauge invariance in the $l$-loop
truncation is due to the missing $\left(l+1\right)$-loop diagrams
(see, e.g. \citep{York,York2012} for a discussion of the gauge fixing
problem).

Several studies have attempted to find a remedy for this problem.
These are discussed in \citep{Pilaftsis2013} and references therein.
Here we restrict attention to the technique most frequently advocated
in the literature \citep{*[{See e.g. }] [] VanHees2002,*Berges2005,*Carrington2009}.
This technique constructs the so called \emph{external propagator}
as the second functional derivative of a resummed effective action
which depends only on the mean field, obtained by eliminating the
2- through $n$-point correlation functions of the $n$PIEA by their
equations of motion. The resulting effective action does obey a 1PI
type Ward identity and the external propagator yields massless Goldstone
bosons. However, the external propagator is not the propagator used
in loop graphs, so the loop corrections still contain massive Goldstone
bosons leading to incorrect thresholds, decay rates and violations
of unitarity. In order to avoid these problems a manifestly self-consistent
scheme must be used.

Pilaftsis and Teresi recently developed a method which circumvents
these difficulties \citep{Pilaftsis2013} for the widely used 2PIEA
(also known as the CJT effective action after Cornwall, Jackiw and
Tomboulis \citep{Cornwall1974}, the Luttinger-Ward functional or
$\Phi$-derivable approximation depending on the context). The idea
is incredibly simple: impose the desired Ward identities directly
on the free correlation functions. This is consistently implemented
by using Lagrange multipliers. The remarkable point is that the resulting
equations of motion can be put into a form that completely eliminates
the Lagrange multiplier field. They achieve this by taking a limit
in which the Lagrange multiplier vanishes from all but one of the
equations of motion, and this remaining equation of motion is replaced
with the constraint to obtain a closed system. We show that this nontrivial
aspect of the procedure generalizes to the 3PIEA. We find that the
generalization requires a careful consideration of the variational
procedure, however, and an infinity of schemes are possible. A new
principle is required to choose between the schemes and we propose
what we call the \emph{d'Alembert formalism} as the appropriate principle
by analogy to the constrained variational problem in mechanics.

We extend the work of Pilaftsis and Teresi to the 3PIEA for three
reasons. First, the 3PIEA is known to be the required starting point
to obtain a self-consistent non-equilibrium kinetic theory of gauge
theories. The accurate calculation of transport coefficients and thermalization
times in gauge theories requires the use of $n$PIEA with $n\geq3$
(see, e.g. \citep{Carrington2009,Smolic2012a,Berges2004} and references
therein for discussion). The fundamental reason for this is that the
3PIEA includes medium induced effects on the three-point vertex at
leading order. The 2PIEA in gauge theory contains a dressed propagator
but not a dressed vertex, leading not only to an inconsistency of
the resulting kinetic equation but also to a spurious gauge dependence
of the kind discussed previously. We consider this work to be a stepping
stone towards a fully self-consistent, nonperturbative and manifestly
gauge invariant treatment of out of equilibrium gauge theories.

Second, $n$PIEA allows one to accurately describe the initial value
problem with 1- to $n$-point connected correlation functions in the
initial state. For example, the widely used 2PIEA allows one to solve
the initial value problem for initial states with a Gaussian density
matrix. However, the physical applications one has in mind typically
start from a near thermal equilibrium state which is not well approximated
by a Gaussian density matrix. This leads to problems with renormalization,
unphysical transient responses and thermalization to the wrong temperature
\citep{Garny2009a}. This is addressed in \citep{Garny2009a,*VanLeeuwen2011,*VanLeeuwen2013}
by the addition of an infinite set of nonlocal vertices which only
have support at the initial time. Going to $n>2$ allows one to better
describe the initial state, thereby reducing the need for additional
nonlocal vertices.

Lastly, the infinite hierarchy of $n$PIEA is the natural home for
the widely used 2PIEA (in all its guises) and provides the clearest
route for systematic improvements over existing treatments. Thus investigating
symmetry improvement of 3PIEA is a well motived next step in the development
of non-perturbative QFT.

After this introductory section we review $n$PIEA in Section \ref{sec:Review-of-nPI},
focusing on the 3PIEA for a model $\mathrm{O}\left(N\right)$ scalar
field theory with symmetry breaking as a specific example. Then in
Section \ref{sec:Symmetry-improvement} we review and extend the symmetry
improvement program of Pilaftsis and Teresi. This includes a derivation
of the required Ward identities, their implementation as constraints
using Lagrange multipliers and the limiting procedure required to
obtain sensible equations of motion for the system. We will see that
this procedure rests on a certain technical assumption which we will
justify in Appendix \ref{sub:The-d'Alembert-formalism} and make a
connection to the d'Alembert\emph{ }principle using a mechanical analogy.
Then in Section \ref{sec:Renormalization} we investigate the renormalization
of the theory, first with the two loop truncation and then three loops.
The three loop truncation is analytically intractable in $1+3$ dimensions
so, after discussing the renormalization procedure in arbitrary dimension,
we present results for $1+2$ dimensions. The result of this section
is a set of finite equations of motion which must be solved numerically.
In Section \ref{sec:Hartree-approximation} we solve the theory at
the Hartree-Fock level and discuss the phase transition thermodynamics.
Section \ref{sub:Two-dimensions-and} is a verification that the Coleman-Mermin-Wagner
theorem holds in the symmetry improvement formalism despite the imposition
of Ward identities, a check on the consistency of the formalism. In
Section \ref{sec:Optical-theorem-and} we discuss the effects of symmetry
improvement on the absorptive parts of propagators and make some comments
involving the Higgs decay rate and dispersion relations. Finally in
Section \ref{sec:Discussion} we discuss the main themes of the paper
and point out directions for future work.

On notation: we work mostly in 1+3 dimensions with $\eta_{\mu\nu}=\mathrm{diag}\left(1,-1,-1,-1\right)$,
although the generalization to other dimensions is simple. We take
$\hbar=c=k_{\mathrm{B}}=1$ as far as units are concerned, though we keep loop counting
factors of $\hbar$ explicit. Repeated indices are summed. Often,
field indices accompany spacetime arguments. Repeated indices in this
case imply an integration over the corresponding spacetime argument
as well (``DeWitt notation''). Where explicitly indicated, spacetime
and momentum integrals are written in compressed notation with $\int_{x}\equiv\int\mathrm{d}^{4}x$,
$\int_{p}\equiv\int\mathrm{d}^{4}p/\left(2\pi\right)^{4}$ and $\int_{\boldsymbol{p}}\equiv\int_{\boldsymbol{p}}\mathrm{d}^{3}\boldsymbol{p}/\left(2\pi\right)^{3}$
etc. $\left\langle \mathrm{T}\left[\cdots\right]\right\rangle $ represents
the time ordered product of the factors in $\left[\cdots\right]$.
Through most of this article the meaning of time ordering is left
implicit. The formalism can be readily applied to vacuum field theory
($t\in\left(-\infty,+\infty\right)$ with the natural ordering), finite
temperature field theory in the imaginary time or Matsubara formalism
($t\to-i\tau$, with periodic boundary conditions on $\tau\in\left[0,\beta=\frac{1}{k_{\mathrm{B}}T}\right)$
and the natural ordering on $\tau$) \citep{Kapusta2006}, and general
non-equilibrium field theory on the two time Schwinger-Keldysh contour
($t$ runs from $0$ to $+\infty$ then from $+\infty-i\epsilon$
back down to $0-i\epsilon$ with time ordering in the sense of position
along the contour rather than the magnitude $\left|t\right|$) \citep{Rammer1986,*Stefanucci2013}.
In Section \ref{sec:Renormalization} we develop the renormalization
theory for the vacuum case and in Section \ref{sec:Hartree-approximation}
we solve the Hartree-Fock approximation at finite temperature in the
Matsubara formalism.

\section{Review of $n$PI effective actions\label{sec:Review-of-nPI}}

For the sake of having an explicit example we consider the $\mathrm{O}\left(N\right)$
linear $\sigma$-model given by the action 
\begin{equation}
S\left[\phi\right]=\int_{x}\frac{1}{2}\partial_{\mu}\phi_{a}\partial^{\mu}\phi^{a}-\frac{1}{2}m^{2}\phi_{a}\phi^{a}-\frac{\lambda}{4!}\left(\phi_{a}\phi^{a}\right)^{2},\label{eq:classical-action}
\end{equation}
where $a=1,\cdots,N$ is the flavor index. In the symmetry breaking
regime $m^{2}<0$ and a vacuum expectation value develops, which by
symmetry can be taken in the last component $\left\langle \phi\right\rangle =\left(0,\ldots,0,v\right)$
where $v^{2}=-6m^{2}/\lambda$ at tree level. The massive mode, which
we loosely call ``the Higgs'' (reflecting our ultimate interest
in the Standard Model, despite the absence of gauge interactions in
\eqref{eq:classical-action}), gets a tree level mass $m_{H}^{2}=\lambda v^{2}/3=-2m^{2}$.

The $n$PI effective actions form a systematic hierarchy of functionals
$\Gamma^{\left(n\right)}\left[\varphi,\Delta,V,\cdots,V^{\left(n\right)}\right]$
where $\varphi,\cdots,V^{\left(n\right)}$ are the proper 1- through
$n$-point correlation functions and we have suppressed spacetime
arguments and flavour indices. In more detail,

\begin{align}
\varphi_{a} & =\left\langle \phi_{a}\right\rangle ,\label{eq:varphi-def}\\
\Delta_{ab} & =i\hbar\left(\left\langle \mathrm{T}\left[\phi_{a}\phi_{b}\right]\right\rangle -\left\langle \phi_{a}\right\rangle \left\langle \phi_{b}\right\rangle \right),\label{eq:Delta_ab-def}\\
\hbar^{2}\Delta_{ad}\Delta_{be}\Delta_{cf}V_{def} & =\left\langle \mathrm{T}\left[\phi_{a}\phi_{b}\phi_{c}\right]\right\rangle -\left\langle \mathrm{T}\left[\phi_{a}\phi_{b}\right]\right\rangle \left\langle \phi_{c}\right\rangle \nonumber \\
 & -\left\langle \mathrm{T}\left[\phi_{c}\phi_{a}\right]\right\rangle \left\langle \phi_{b}\right\rangle -\left\langle \mathrm{T}\left[\phi_{b}\phi_{c}\right]\right\rangle \left\langle \phi_{a}\right\rangle \nonumber \\
 & +2\left\langle \phi_{a}\right\rangle \left\langle \phi_{b}\right\rangle \left\langle \phi_{c}\right\rangle \label{eq:V3-def}\\
\vdots\nonumber 
\end{align}
In general $V^{\left(n\right)}$ is the sum of connected one particle
irreducible Feynman diagrams contributing to $\left\langle \phi^{n}\right\rangle $
with all external legs (including leg corrections) removed.

In the absence of external source terms the correlation functions
obey equations of motion of the form
\begin{equation}
\frac{\delta\Gamma^{\left(n\right)}}{\delta\varphi}=0,\ \frac{\delta\Gamma^{\left(n\right)}}{\delta\Delta}=0,\ \cdots,\ \frac{\delta\Gamma^{\left(n\right)}}{\delta V^{\left(n\right)}}=0.\label{eq:npi-eom}
\end{equation}
In the exact theory $\Gamma^{\left(n\right)}$ obey equivalence relationships
\begin{equation}
\Gamma^{\left(1\right)}\left[\varphi\right]=\Gamma^{\left(2\right)}\left[\varphi,\Delta\right]=\Gamma^{\left(3\right)}\left[\varphi,\Delta,V\right]=\cdots,\label{eq:npi-equivalence}
\end{equation}
where extra arguments are eliminated by their equations of motion
when comparisons are made. These relationships only hold approximately
when approximations are made to the theory. A stronger equivalence
hierarchy that relates loop-wise truncations of the $\Gamma^{\left(n\right)}$
will be discussed below.

For later convenience we introduce the tree level vertex functions
\begin{align}
V_{0abc}\left(x,y,z\right) & =\left.\frac{\delta^{3}S\left[\phi\right]}{\delta\phi_{a}\left(x\right)\delta\phi_{b}\left(y\right)\delta\phi_{c}\left(z\right)}\right|_{\phi=\varphi},\\
W_{abcd}\left(x,y,z,w\right) & =\left.\frac{\delta^{4}S\left[\phi\right]}{\delta\phi_{a}\left(x\right)\delta\phi_{b}\left(y\right)\delta\phi_{c}\left(z\right)\delta\phi_{d}\left(w\right)}\right|_{\phi=\varphi}.
\end{align}
For the $\mathrm{O}\left(N\right)$ model these are
\begin{align}
V_{0abc}\left(x,y,z\right) & =-\frac{\lambda}{3}\left[\delta_{ab}\varphi_{c}\left(x\right)+\delta_{ca}\varphi_{b}\left(x\right)+\delta_{bc}\varphi_{a}\left(x\right)\right]\nonumber \\
 & \times\delta^{\left(4\right)}\left(x-y\right)\delta^{\left(4\right)}\left(x-z\right),\\
W_{abcd}\left(x,y,z,w\right) & =-\frac{\lambda}{3}\left[\delta_{ab}\delta_{cd}+\delta_{ac}\delta_{bd}+\delta_{ad}\delta_{bc}\right]\nonumber \\
 & \times\delta^{\left(4\right)}\left(x-y\right)\delta^{\left(4\right)}\left(x-z\right)\delta^{\left(4\right)}\left(x-w\right).
\end{align}

The $n$PIEA is defined in the functional integral formalism by the
Legendre transform of the connected generating function 
\begin{equation}
W^{\left(n\right)}\left[J,K^{\left(2\right)},\cdots,K^{\left(n\right)}\right]=-i\hbar\ln Z^{\left(n\right)}\left[J,K^{\left(2\right)},\cdots,K^{\left(n\right)}\right],
\end{equation}
for a field theory in the presence of source terms defined by the
generating functional
\begin{widetext}
\begin{equation}
Z^{\left(n\right)}\left[J,K^{\left(2\right)},\cdots,K^{\left(n\right)}\right]=\int\mathcal{D}\left[\phi\right]\exp\frac{i}{\hbar}\left(S\left[\phi\right]+J_{x}\phi_{x}+\frac{1}{2}\phi_{x}K_{xy}^{\left(2\right)}\phi_{y}+\cdots+\frac{1}{n!}K_{x_{1}\cdots x_{n}}^{\left(n\right)}\phi_{x_{1}}\cdots\phi_{x_{n}}\right).\label{eq:npi-z-functional}
\end{equation}

\end{widetext}

Then $\Gamma^{\left(n\right)}$ is the $n$-fold Legendre transform
\begin{align}
\Gamma^{\left(n\right)}\left[\varphi,\Delta,V,\cdots,V^{\left(n\right)}\right] & =W^{\left(n\right)}-J\frac{\delta W^{\left(n\right)}}{\delta J}\nonumber \\
 & -K^{\left(2\right)}\frac{\delta W^{\left(n\right)}}{\delta K^{\left(2\right)}}-\cdots\nonumber \\
 & -K^{\left(n\right)}\frac{\delta W^{\left(n\right)}}{\delta K^{\left(n\right)}},\label{eq:npi-legendre-transform}
\end{align}
where the source terms $J,K^{\left(2\right)},\cdots,K^{\left(n\right)}$
are solved for in terms of the $\varphi,\Delta,\cdots,V^{\left(n\right)}$.
Spacetime integrations and $\mathrm{O}\left(N\right)$ index contractions
have been suppressed for brevity. For bosonic fields the $\Delta,\cdots,V^{\left(n\right)}$
are totally symmetric under permutations of their arguments. The generalisation
to fermions requires sign changes for odd permutations of arguments
corresponding to fermionic fields, but is otherwise straightforward.
(Note that the $n$PIEA is defined by this Legendre transform, \emph{not}
by any irreducibility property of the Feynman graphs, though for low
enough loop orders the graphs are irreducible as the name implies.
At high enough loop order for $n>2$ the name becomes misleading.
For example, the five loop 5PIEA contains graphs that are not five-particle
irreducible \citep{Carrington2011a}!)

$\Gamma^{\left(1\right)}\left[\varphi\right]$ is the familiar 1PI
effective action introduced by Goldstone, Salam and Weinberg and 
independently by Jona-Lasinio \citep{Goldstone1962,*Jona-Lasinio1964}.
$\Gamma^{\left(1\right)}\left[\varphi\right]$
can be written
\begin{equation}
\Gamma^{\left(1\right)}\left[\varphi\right]=S\left[\varphi\right]+\frac{i\hbar}{2}\mathrm{Tr}\ln\left\{ \Delta_{0}^{-1}\left[\varphi\right]\right\} +\Gamma_{2}^{\left(1\right)}\left[\varphi\right],\label{eq:1piea}
\end{equation}
where $\Delta_{0}^{-1}\left[\varphi\right]=\left.\delta^{2}S\left[\phi+\varphi\right]/\delta\phi^{2}\right|_{\phi=0}$
is the inverse propagator and $\Gamma_{2}^{\left(1\right)}$ is the
sum of all connected vacuum graphs with $\geq2$ loops where the propagators
$\Delta_{0}\left[\varphi\right]$ and vertices are obtained from the
shifted action $S\left[\phi+\varphi\right]$ with the additional prescription
that all 1-particle reducible graphs are dropped \citep{Zinn-Justin1990}.
Note that \eqref{eq:V3-def} is equivalent to
\begin{equation}
V^{\left(1\right)}_{def}\left(u,v,w\right)=\frac{\delta^{3}\Gamma^{\left(1\right)}}{\delta\varphi_{d}\left(u\right)\delta\varphi_{e}\left(v\right)\delta\varphi_{f}\left(w\right)},\label{eq:3pt-vertex-in-terms-of-gamma1}
\end{equation}
where the superscript ``$\left(1\right)$'' indicates the vertex derived
from the 1PIEA.

The 2PIEA $\Gamma^{\left(2\right)}\left[\varphi,\Delta\right]$ was
introduced in the context of non-relativistic statistical mechanics,
apparently independently, by Lee and Yang, Luttinger and Ward, and
others, but was brought to the functional formalism and relativistic
field theory by Cornwall, Jackiw and Tomboulis \citep{Cornwall1974}.
$\Gamma^{\left(2\right)}\left[\varphi,\Delta\right]$ is most easily
computed by noting that the Legendre transform can be performed in
stages. First perform the Legendre transform with respect to $J$,
using result \eqref{eq:1piea} with the replacement $S\left[\phi\right]\to S\left[\phi\right]+\frac{1}{2}\phi_{x}K_{xy}^{\left(2\right)}\phi_{y},$
then do the transform with respect to $K^{\left(2\right)}$. This
procedure leads to (up to an irrelevant constant)
\begin{align}
\Gamma^{\left(2\right)}\left[\varphi,\Delta\right] &= S\left[\varphi\right]+\frac{i\hbar}{2}\mathrm{Tr}\ln\left(\Delta^{-1}\right)+\frac{i\hbar}{2}\mathrm{Tr}\left(\Delta_{0}^{-1}\Delta\right) \nonumber\\
 & +\Gamma_{2}^{\left(2\right)}\left[\varphi,\Delta\right].\label{eq:2piea}
\end{align}

The equation of motion for $\Delta$ is Dyson's equation,

\begin{equation}
\Delta^{-1}=\Delta_{0}^{-1}-\Sigma,\label{eq:dyson-eq}
\end{equation}
where
\begin{equation}
\Sigma=\frac{2i}{\hbar}\frac{\delta\Gamma_{2}^{\left(2\right)}\left[\varphi,\Delta\right]}{\delta\Delta},\label{eq:generic-self-energy}
\end{equation}
is identified as the 1PI self-energy. Since $\Sigma$ consists of
1PI two-point graphs, $\Gamma_{2}^{\left(2\right)}$ must consist
of 2PI vacuum graphs. That is, $\Gamma_{2}^{\left(2\right)}$ is the
sum of all vacuum diagrams which do not fall apart when any two lines
are cut. This results in a drastic reduction in the number of graphs
at a given loop order. Further, the propagators in a 2PI graph are
the full propagators $\Delta$, with all self-energy insertions resummed
to infinite order.

The 3PIEA $\Gamma^{\left(3\right)}\left[\varphi,\Delta,V\right]$
can be computed following the same method: replace $S\left[\phi\right]\to S\left[\phi\right]+\frac{1}{3!}K_{xyz}^{\left(3\right)}\phi_{x}\phi_{y}\phi_{z}$
in the previous result and perform the Legendre transform with respect
to $K^{\left(3\right)}$.

\begin{figure}
\includegraphics[width=1\columnwidth]{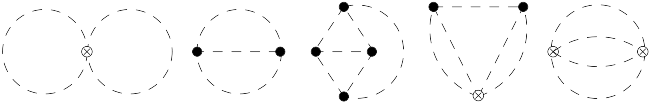}

\protect\caption{\label{fig:2pi-diags-up-to-3-loops}Two and three loop diagrams contributing
to $\Gamma_{3}^{\left(3\right)}\left[\varphi,\Delta,V\right]$.
We label these $\Phi_{1}$ through $\Phi_{5}$ from left to right,
respectively, and their explicit forms are given in \eqref{eq:phi1}
through \eqref{eq:phi5}. Solid circles represent the resummed vertices
$V$ and the crossed circles represent the bare vertices
$V_{0}$ and $W$. The dashed
lines represent the resummed propagators $\Delta$. Note that these
diagrams are called ``\noun{eight},'' ``\noun{egg},'' ``\noun{mercedes},''
``\noun{hair},'' and ``\noun{bball}'' respectively in the nomenclature
of \citep{Carrington2011a,Carrington2011}.}
\end{figure}

The shift by the source term $K^{\left(3\right)}$ results in the
introduction of an effective three point vertex $\tilde{V}\equiv V_{0}+K^{\left(3\right)}$
appearing in $\Gamma_{2}^{\left(2\right)}$. The difficult step of
the Legendre transform is relating $\tilde{V}$ to $V$.
This can be done by comparing $\delta W^{\left(3\right)}\left[J,K^{\left(2\right)},K^{\left(3\right)}\right]/\delta K^{\left(3\right)}$
with $\delta\Gamma^{\left(2\right)}\left[\varphi,\Delta;\tilde{V}\right]/\delta K^{\left(3\right)}$
(see \citep{Berges2004,Berges2004a}). The final result for $\Gamma^{\left(3\right)}$ is
\begin{equation}
\Gamma^{\left(3\right)}=S\left[\varphi\right]+\frac{i\hbar}{2}\mathrm{Tr}\ln\left(\Delta^{-1}\right)+\frac{i\hbar}{2}\mathrm{Tr}\left(\Delta_{0}^{-1}\Delta\right)+\Gamma_{3}^{\left(3\right)},\label{eq:3piea}
\end{equation}
where to three loop order the diagram piece is
\begin{align}
\Gamma_{3}^{\left(3\right)} & =\Phi_{1}+\frac{\hbar^{2}}{3!}V_{0}\Delta\Delta\Delta V-\Phi_{2}\nonumber \\
 & +\Phi_{3}+\Phi_{4}+\Phi_{5}+\mathcal{O}\left(\hbar^{4}\right),
\end{align}
where $\Phi_{1},\cdots,\Phi_{5}$ are given by the Feynman diagrams
shown in Figure \ref{fig:2pi-diags-up-to-3-loops}. Explicitly,

\begin{align}
\Phi_{1} & =-\frac{\hbar^{2}}{8}W_{abcd}\Delta_{ab}\Delta_{cd},\label{eq:phi1}\\
\Phi_{2} & =\frac{\hbar^{2}}{12}V_{abc}V_{def}\Delta_{ad}\Delta_{be}\Delta_{cf},\label{eq:phi2}\\
\Phi_{3} & =\frac{i\hbar^{3}}{4!}V_{abc}V_{def}V_{ghi}V_{jkl}\Delta_{ad}\Delta_{bg}\Delta_{cj}\Delta_{eh}\Delta_{fk}\Delta_{il},\label{eq:phi3}\\
\Phi_{4} & =-\frac{i\hbar^{3}}{8}V_{abc}V_{def}W_{ghij}\Delta_{ad}\Delta_{bg}\Delta_{ch}\Delta_{ei}\Delta_{fj},\label{eq:phi4}\\
\Phi_{5} & =\frac{i\hbar^{3}}{48}W_{abcd}W_{efgh}\Delta_{ae}\Delta_{bf}\Delta_{cg}\Delta_{dh}.\label{eq:phi5}
\end{align}

\begin{figure}
\includegraphics[width=1\columnwidth]{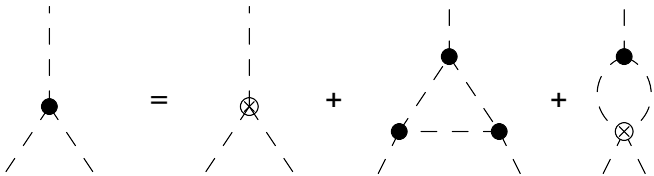}\protect\caption{\label{fig:vertex-eom}Equation of motion for the 3PI vertex function
$V$ up to one loop order \eqref{eq:3pi-vertex-eom}.
Note that the bubble graph (last term) is implicitly symmetrized over
external momenta and $\mathrm{O}\left(N\right)$ indices.}
\end{figure}

The 3PI equation of motion for $V$ is $0=\frac{\delta\Gamma^{\left(3\right)}}{\delta V_{abc}}$,
which reads in full
\begin{align}
V_{abc} & =V_{0abc}+i\hbar V_{ade}V_{bfg}V_{chi}\Delta_{df}\Delta_{eh}\Delta_{gi}\nonumber \\
 & -\frac{1}{3!}\sum_{\pi}\frac{3i\hbar}{2}V_{\pi\left(a\right)de}W_{\pi\left(b\right)\pi\left(c\right)fg}\Delta_{df}\Delta_{eg}+\mathcal{O}\left(\hbar^{2}\right),\label{eq:3pi-vertex-eom}
\end{align}
where $\sum_{\pi}$ is a sum over the $3!$ permutations mapping $\left(a,b,c\right)\to\left(\pi\left(a\right),\pi\left(b\right),\pi\left(c\right)\right)$
(spacetime arguments are permuted as well). The graphical interpretation
of this equation is shown in Figure \ref{fig:vertex-eom}. The permutations
lead to the usual $s$, $t$, and $u$ channel contributions with the expected
symmetry factors. This equation is best thought of as a self-consistent integral
equation in the same spirit as a Schwinger-Dyson equation, and can be solved
iteratively. By iterating \eqref{eq:3pi-vertex-eom} one sees that it sums a
sequence of vertex correction diagrams to infinite order.

If all higher order terms are kept in $\Gamma_{3}^{\left(3\right)}$ the
resulting $V$ is the same as the $V^{\left(1\right)}$ of
\eqref{eq:3pt-vertex-in-terms-of-gamma1}, however truncated actions give
solutions $V \neq V^{\left(1\right)}$. Similar remarks apply for the propagators.
These self-consistent solutions do not in general obey the desirable field
theoretic properties of the full solution, such as Ward identities.
The symmetry improvement strategy is to impose 1PI Ward identities which as
constraints on the self-consistent solutions $\Delta$ and $V$. This is discussed
further in Section \ref{sec:Symmetry-improvement}.

Note that \eqref{eq:3pi-vertex-eom} can be derived by removing a
resummed vertex from each graph in $\Gamma_{3}^{\left(3\right)}$
(because $\delta/\delta V$ acts by removing a single
$V$ factor from graphs in all possible ways), which
has the graphical effect of opening two loops. This means that the
one loop correction to $V$ comes from three loop
graphs in $\Gamma^{\left(3\right)}$. Thus a loop-wise truncation
of $n$PIEA for $n\geq3$ does not lead to a loop-wise truncation
of the corresponding equations of motion. We will discuss the further
implications of this in Section \ref{sec:Optical-theorem-and}.

Another important implication of this result is that $\Gamma^{\left(2\right)}$
and $\Gamma^{\left(3\right)}$ are equivalent to two loop order (after
one substitutes $V=V_{0}+\mathcal{O}\left(\hbar\right)$
in $\Gamma^{\left(3\right)}$). However, $\Gamma^{\left(2\right)}$
and $\Gamma^{\left(3\right)}$ differ at three loop order because
$\Gamma^{\left(3\right)}$ contains resummed vertex corrections that
$\Gamma^{\left(2\right)}$ does not. This is an example of an equivalence
hierarchy of $n$PI effective actions that has the general form \citep{Berges2004}:
\begin{align}
\Gamma_{\left(\text{1 loop}\right)}^{\left(1\right)}\left[\phi\right] & =\Gamma_{\left(\text{1 loop}\right)}^{\left(2\right)}\left[\phi,\Delta\right]=\cdots,\\
\Gamma_{\left(\text{2 loop}\right)}^{\left(1\right)}\left[\phi\right] & \neq\Gamma_{\left(\text{2 loop}\right)}^{\left(2\right)}\left[\phi,\Delta\right]=\Gamma_{\left(\text{2 loop}\right)}^{\left(3\right)}\left[\phi,\Delta,V\right]=\cdots,\\
\Gamma_{\left(\text{3 loop}\right)}^{\left(1\right)}\left[\phi\right] & \neq\Gamma_{\left(\text{3 loop}\right)}^{\left(2\right)}\left[\phi,\Delta\right]\neq\Gamma_{\left(\text{3 loop}\right)}^{\left(3\right)}\left[\phi,\Delta,V\right]\nonumber \\
 & =\Gamma_{\left(\text{3 loop}\right)}^{\left(4\right)}\left[\phi,\Delta,V,V^{\left(4\right)}\right]=\cdots,\\
\vdots\nonumber 
\end{align}
where the subscripts represent the order of the loop-wise truncation
and the ``extra'' correlation functions are to be evaluated at the
solutions of their respective equations of motion before making the
comparison (and also allowance is made for shifts by irrelevant constants).
This equivalence hierarchy has been explicitly checked up to five
loop 5PI order in scalar field theories \citep{Carrington2011a}.

The existence of the equivalence hierarchy implies that in the standard
formalism one gains nothing by going to higher $n$PI effective actions
unless one also includes diagrams with at least $n$ loops, since
for $m>n$ one can always reduce $\Gamma_{\left(n\text{ loop}\right)}^{\left(m\right)}$
to $\Gamma_{\left(n\text{ loop}\right)}^{\left(n\right)}$. However,
we shall see that symmetry improvement breaks this equivalence hierarchy.
In particular, we find that the symmetry improvement of the 3PI effective
action modifies the $\Delta$ equation of motion in a way that remains
non-trivial even if $\Gamma^{\left(3\right)}$ is then truncated at
two loops and $V$ is replaced by its tree level
value $V_{0}$. In general we find that the symmetry
improvement procedure introduces Ward identities that relate $k$-point
functions to $\left(k+1\right)$-point functions and these constraints
spoil the equivalence hierarchy, i.e. the ``operations'' of symmetry
improvement and reduction in the hierarchy do not commute. The consequences
of this for the phase diagram of the scalar $\mathrm{O}\left(N\right)$
theory in the various possible schemes are investigated in Section
\eqref{sec:Hartree-approximation}.

\section{Symmetry improvement\label{sec:Symmetry-improvement}}

Symmetry improvement begins with the consideration of the Ward identities
in the $n$PI formalism. Following \citep{Pilaftsis2013} we derive
the Ward identities from the condition that the effective action is
invariant under a symmetry transformation. The theory in \eqref{eq:classical-action}
has the $\mathrm{O}\left(N\right)$ symmetry transform
\begin{equation}
\phi_{a}\to\phi_{a}+i\epsilon_{A}T_{ab}^{A}\phi_{b},\label{eq:O(N)-transform}
\end{equation}
where $T^{A}$ are the generators of the group in the fundamental
representation ($A=1,\cdots,N\left(N-1\right)/2$) and $\epsilon_{A}$
are infinitesimal transformation parameters. Note that our implicit
integration convention can be maintained if we consider that $T_{ab}^{A}$
contains a spacetime delta function $T_{ab}^{A}\propto\delta^{\left(4\right)}\left(x_{a}-x_{b}\right)$.
Also $T_{ab}^{A}=-T_{ba}^{A}$. Under this transformation the effective
actions change by
\begin{widetext}
\begin{align}
\delta\Gamma^{\left(1\right)} & =\frac{\delta\Gamma^{\left(1\right)}}{\delta\varphi_{a}}i\epsilon_{A}T_{ab}^{A}\varphi_{b},\label{eq:1pi-transform}\\
\delta\Gamma^{\left(2\right)} & =\frac{\delta\Gamma^{\left(2\right)}}{\delta\varphi_{a}}i\epsilon_{A}T_{ab}^{A}\varphi_{b}+\frac{\delta\Gamma^{\left(2\right)}}{\delta\Delta_{ab}}i\epsilon_{A}\left(T_{ac}^{A}\Delta_{cb}+T_{bc}^{A}\Delta_{ac}\right),\label{eq:2pi-transform}\\
\delta\Gamma^{\left(3\right)} & =\frac{\delta\Gamma^{\left(3\right)}}{\delta\varphi_{a}}i\epsilon_{A}T_{ab}^{A}\varphi_{b}+\frac{\delta\Gamma^{\left(3\right)}}{\delta\Delta_{ab}}i\epsilon_{A}\left(T_{ac}^{A}\Delta_{cb}+T_{bc}^{A}\Delta_{ac}\right)+\frac{\delta\Gamma^{\left(3\right)}}{\delta V_{abc}}i\epsilon_{A}\left(T_{ad}^{A}V_{dbc}+T_{bd}^{A}V_{adc}+T_{cd}^{A}V_{abd}\right),\label{eq:3pi-transform}\\
\vdots\nonumber 
\end{align}

\end{widetext}
according to the tensorial structure of the arguments. The next steps
to derive the Ward identities are to set $\delta\Gamma^{\left(n\right)}=0$,
take functional derivatives of the resulting equations with respect
to $\varphi$ and finally apply the equations of motion. We also extract
the overall factors of $i\epsilon_{A}$. We call the identity derived
from the $m$-th derivative of $\delta\Gamma^{\left(n\right)}$ the
$\left(m+1\right)$-point $n$PI Ward identity, denoted by $\mathcal{W}_{a_{1}\cdots a_{m}}^{A\left(n\right)}=0$
where $a_{1},\cdots,a_{m}$ are $\mathrm{O}\left(N\right)$/spacetime
indices. We note first of all that $\mathcal{W}^{\left(n\right)}=0$
identically by the equations of motion. We also find that
\begin{align}
\mathcal{W}_{c}^{A\left(1\right)} & =\frac{\delta\Gamma^{\left(1\right)}}{\delta\varphi_{c}\delta\varphi_{a}}T_{ab}^{A}\varphi_{b},\label{eq:2pt-1pi-WI-generic}\\
\mathcal{W}_{cd}^{A\left(1\right)} & =\frac{\delta\Gamma^{\left(1\right)}}{\delta\varphi_{d}\delta\varphi_{c}\delta\varphi_{a}}T_{ab}^{A}\varphi_{b}+\frac{\delta\Gamma^{\left(1\right)}}{\delta\varphi_{c}\delta\varphi_{a}}T_{ad}^{A}+\frac{\delta\Gamma^{\left(1\right)}}{\delta\varphi_{d}\delta\varphi_{a}}T_{ac}^{A}.\label{eq:3pt-1pi-WI-generic}
\end{align}

Specialising now to the broken symmetry vacuum $\varphi_{b}=v\delta_{bN}$
we obtain the following identities by substituting different generators
$T_{ab}^{A}$ in turn:
\begin{align}
0 & =\int_{x_{a}}\Delta_{ca}^{-1}\left(x_{c},x_{a}\right)v,\ a\neq N,\label{eq:w1-WI}\\
0 & =\int_{z}V_{Nab}\left(x,y,z\right)v+\Delta_{ab}^{-1}\left(x,y\right)\nonumber\\
  & -\delta_{ab}\Delta_{NN}^{-1}\left(x,y\right),\ a,b\neq N\label{eq:w2-WI}\\
0 & =\Delta_{ca}^{-1},\ a\neq c,\label{eq:w3-WI}\\
0 & =\int_{z}V_{dca}\left(x,y,z\right)v,\ d,c,a\neq N,\label{eq:w4-WI}\\
0 & =\int_{z}V_{NNa}\left(x,y,z\right)v,\ a\neq N.\label{eq:w5-WI}
\end{align}
Note that we explicitly write spacetime arguments, $O\left(N\right)$ indices and
integrations in the above. This is because DeWitt notation would lead to ambiguities here.
Below we introduce an ansatz adapted to the situation which again allows for notational
simplifications.

The essence of symmetry improvement is to impose these Ward identities,
derived for the 1PI correlation functions, on the $n$PI correlation functions.
Effectively, we change
$f\left(\Delta_{\text{1PI}},V_{\text{1PI}}\right) \to f\left(\Delta_{\text{3PI}},V_{\text{3PI}}\right),$
where $f$ is the Ward identity and we change the arguments but not the functional form.
We have already made this substitution in \eqref{eq:w1-WI}-\eqref{eq:w5-WI}.

The first two identities will prove important in the following, however,
the identities \eqref{eq:w3-WI}-\eqref{eq:w5-WI} are trivial in
the sense that they can be satisfied simply by postulating an ansatz
for $\Delta$ and $V$ which is tensorial under the
unbroken $\mathrm{O}\left(N-1\right)$ symmetry. For later convenience
we adopt this spontaneous symmetry breaking (SSB) ansatz now by introducing
the notation

\begin{equation}
\Delta_{ab}\left(x,y\right)=\begin{cases}
\Delta_{G}\left(x,y\right), & a=b\neq N,\\
\Delta_{H}\left(x,y\right), & a=b=N,\\
0, & \text{otherwise},
\end{cases}\label{eq:ssb-propagator-ansatz}
\end{equation}
for the Goldstone $\left(\Delta_{G}\right)$ and Higgs $\left(\Delta_{H}\right)$
propagators respectively, and we also introduce the vertex functions
$\bar{V}$ and $V_{N}$ where

\begin{align}
V_{abc}\left(x,y,z\right) & =\begin{cases}
\bar{V}\left(x,y,z\right)\delta_{aN}\delta_{bc} & \text{exactly one of }\\
+\text{cyclic permutations} & a,b,c=N,\\
\\
V_{N}\left(x,y,z\right) & a=b=c=N,\\
0 & \text{otherwise}.
\end{cases}\label{eq:ssb-vertex-ansatz}
\end{align}

Note that $V_{N}$ is not constrained by any of the 2- or 3-point
Ward identities. Spacetime arguments are permuted along with the $\mathrm{O}\left(N\right)$
indices, so that the first spacetime argument of $\bar{V}$ is always
the one referring to the Higgs. The other two arguments refer to the
Goldstone bosons and $\bar{V}$ is symmetric under their interchange.
$V_{N}$ is totally symmetric in its arguments. For reference note
that at the 2-loop truncation, $V=V_{0}$
and we obtain $\bar{V}=\left(-\lambda v/3\right)\times\delta^{\left(4\right)}\left(x-y\right)\delta^{\left(4\right)}\left(x-z\right)$
and $V_{N}=3\bar{V}$. After substituting the ansatz the diagrams
$\Phi_{1},\cdots,\Phi_{5}$ can be put into the form
\begin{widetext}
\begin{align}
\Phi_{1} & =\frac{\hbar^{2}\lambda}{24}\left(N^{2}-1\right)\Delta_{G}\Delta_{G}+\frac{\hbar^{2}\lambda}{12}\left(N-1\right)\Delta_{G}\Delta_{H}+\frac{\hbar^{2}\lambda}{8}\Delta_{H}\Delta_{H},\label{eq:Phi_1-with-ansatz}\\
\Phi_{2} & =\frac{\hbar^{2}}{4}\left(N-1\right)\bar{V}\bar{V}\Delta_{H}\Delta_{G}\Delta_{G}+\frac{\hbar^{2}}{12}V_{N}V_{N}\left(\Delta_{H}\right)^{3},\label{eq:Phi_2-with-ansatz}\\
\Phi_{3} & =\left(N-1\right)\frac{i\hbar^{3}}{3!}V_{N}\left(\bar{V}\right)^{3}\left(\Delta_{H}\right)^{3}\left(\Delta_{G}\right)^{3}+\frac{i\hbar^{3}}{4!}\left(V_{N}\right)^{4}\left(\Delta_{H}\right)^{6}+\left(N-1\right)\frac{i\hbar^{3}}{8}\left(\bar{V}\right)^{4}\Delta_{H}\Delta_{H}\left(\Delta_{G}\right)^{4},\label{eq:Phi_3-with-ansatz}\\
\Phi_{4} & =\frac{i\hbar^{3}\lambda}{24}\left[2\left(N-1\right)\bar{V}V_{N}\left(\Delta_{H}\right)^{3}\Delta_{G}\Delta_{G}\right.+\left(N^{2}-1\right)\bar{V}\bar{V}\Delta_{H}\left(\Delta_{G}\right)^{4}+3V_{N}V_{N}\left(\Delta_{H}\right)^{5}\nonumber \\
 & \left.+2^{2}\left(N-1\right)\bar{V}\bar{V}\left(\Delta_{G}\right)^{3}\Delta_{H}\Delta_{H}\right],\label{eq:Phi_4-with-ansatz}\\
\Phi_{5} & =\frac{i\hbar^{3}\lambda^{2}}{144}\left\{ \left[\left(N-1\right)\Delta_{G}\Delta_{G}+\Delta_{H}\Delta_{H}\right]^{2}+2\left(N-1\right)\left(\Delta_{G}\right)^{4}+2\left(\Delta_{H}\right)^{4}\right\} .\label{eq:Phi_5-with-ansatz}
\end{align}
\end{widetext}
The suppressed spacetime integrations can be restored by comparing
these expression to the diagrams and using the fact that $\bar{V}$
vertices join one Higgs and two Goldstone lines, while $V_{N}$ vertices
join three Higgs lines. (These expressions can be checked using the
supplemental \noun{Mathematica} notebook \citep{supp}.)

In terms of the SSB ansatz variables the nontrivial Ward identities
are $O\left(N\right)$-scalar equations which read
\begin{align}
0 & =\mathcal{W}_{1}\equiv\int_{z}\Delta_{G}^{-1}\left(x,z\right)v,\label{eq:w1-WI-1}\\
0 & =\mathcal{W}_{2}\equiv\int_{z}\bar{V}\left(x,y,z\right)v+\Delta_{G}^{-1}\left(x,y\right)-\Delta_{H}^{-1}\left(x,y\right).\label{eq:w2-WI-1}
\end{align}
The physical meaning of these can be seen by assuming translation
invariance, substituting 
\begin{align}
\Delta_{G/H}^{-1}\left(p\right) & \equiv\int_{x-y}\mathrm{e}^{ip\cdot\left(x-y\right)}\Delta_{G/H}^{-1}\left(x,y\right)\nonumber \\
 & =p^{2}-m_{G/H}^{2}-\Sigma_{G/H}\left(p\right),
\end{align}
and $\bar{V}=\left(-\lambda v/3\right)\times\delta^{\left(4\right)}\left(x-y\right)\delta^{\left(4\right)}\left(x-z\right)+\delta\bar{V}$
where $\delta\bar{V}$ represents the loop corrections to the vertex.
Matching powers of $\hbar$ (which are implicit in $\Sigma_{G/H}$
and $\delta\bar{V}$) results in
\begin{align}
vm_{G}^{2} & =0,\\
-\frac{\lambda v^{2}}{3}+m_{H}^{2}-m_{G}^{2} & =0,\\
\delta\bar{V}\left(p,-p,0\right)v+\Sigma_{H}\left(p\right)-\Sigma_{G}\left(p\right) & =0,\label{eq:wi-for-the-self-energies}
\end{align}
which are Goldstone's theorem, the tree level relation between the
particle masses, and a relation between the vertex correction (with
one external Goldstone boson leg set to zero momentum) and the self-energies
of the Higgs and Goldstone bosons respectively. The imaginary part
of this last identity can be used to extract a relation between the
Higgs decay rate and the off-shell Goldstone boson self-energy and
vertex corrections. This will be investigated in Section \ref{sec:Optical-theorem-and}.

We now wish to impose \eqref{eq:w1-WI-1}-\eqref{eq:w2-WI-1} as constraints
on the allowable values of $\varphi$, $\Delta$ and $V$
in the 3PIEA. First we review the 2PIEA case as discussed in \citep{Pilaftsis2013},
which imposes \eqref{eq:w1-WI-1} on $\Gamma^{\left(2\right)}$ through
the introduction of Lagrange multiplier fields $\ell_{A}^{d}\left(x\right)$,
where $A$ is an $\mathrm{O}\left(N\right)$ adjoint index, and the
symmetry improved effective action which we write in manifestly covariant
form as
\begin{equation}
\tilde{\Gamma}\left[\varphi,\Delta,\ell\right]=\Gamma^{\left(2\right)}\left[\varphi,\Delta\right]+\frac{i}{2}\int_{x}\ell_{A}^{d}\left(x\right)\mathcal{W}_{c}^{A\left(1\right)}\left(x\right)\left[P_{T}\left(\varphi,x\right)\right]_{cd}.
\end{equation}
The transverse projector
\begin{equation}
\left[P_{T}\left(\varphi,x\right)\right]_{cd}=\delta_{cd}-\frac{\varphi_{c}\left(x\right)\varphi_{d}\left(x\right)}{\varphi^{2}\left(x\right)},
\end{equation}
ensures that only the Goldstone modes are involved in the constraint.
The equations of motion follow from $\delta\tilde{\Gamma}/\delta\varphi=\delta\tilde{\Gamma}/\delta\Delta=0$.

Substituting the SSB ansatz and using translation invariance gives
\begin{equation}
\tilde{\Gamma}\left[\varphi,\Delta,\ell\right]=\Gamma^{\left(2\right)}\left[\varphi,\Delta\right]-\ell\mathcal{W}_{1},
\end{equation}
where we have absorbed group theory factors in $\ell$. The 2PI equations
of motion become
\begin{align}
\frac{\partial\Gamma^{\left(2\right)}/VT}{\partial v} & =\ell\frac{\partial}{\partial v}\mathcal{W}_{1},\\
\frac{\delta\Gamma^{\left(2\right)}}{\delta\Delta_{G}\left(x,y\right)} & =-v\ell\left[\int_{x}\Delta_{G}^{-1}\left(x,0\right)\right]^{2},\\
\frac{\delta\Gamma^{\left(2\right)}}{\delta\Delta_{H}\left(x,y\right)} & =0,\\
0 & =v\int_{y}\Delta_{G}^{-1}\left(x,y\right).
\end{align}
The factor of $VT$ on the left hand side of the first equation is
the volume of spacetime, which we have divided by to give an intensive
quantity.

Now applying the constraint with $v\neq0$ directly in the equations
of motion would give zero right hand sides, reducing to the standard
2PI formalism. This is valid in the full theory because the Ward identity
is satisfied. However, this is impossible in the case where the 2PI
effective action is truncated at finite loop order because the actual
Ward identity obeyed by the 2PIEA is $\mathcal{W}_{c}^{A\left(2\right)}\neq\mathcal{W}_{c}^{A\left(1\right)}$.
The manifestation of this fact in the symmetry improvement formalism
is a singularity: $\ell\to\infty$ as $v\int\Delta_{G}^{-1}\to0$
so as to leave a finite right hand side in the first equation of motion.

It is now necessary to introduce the constraint through a limit process,
and choose the scaling of $\ell$ in the limit such that the scalar
equation of motion is traded for the constraint. To this end we set
\begin{equation}
v\int_{y}\Delta_{G}^{-1}\left(x,y\right)=\eta m^{3},\label{eq:regulated-2pi-WI}
\end{equation}
and take the limit $\eta\to0$. Note that, in extension of \citet{Pilaftsis2013},
one may allow separate regulators $\eta_{i}$ for each Goldstone mode
$i=1,\ldots,N-1$, but there is nothing much to gain from this and
it leads to no new difficulties so we take a common regulator $\eta_{i}=\eta$.
$m$ is an arbitrary fixed mass scale, conveniently taken to be $\sim m_{H}$,
which serves to make $\eta$ dimensionless. The modified equations
of motion become
\begin{align}
\frac{\partial\Gamma^{\left(2\right)}/VT}{\partial v} & =\frac{\ell\eta}{v}m^{3},\\
\frac{\delta\Gamma^{\left(2\right)}}{\delta\Delta_{G}\left(z,w\right)} & =-\frac{\ell\eta^{2}}{v}m^{6}.
\end{align}
If we choose to scale $\eta$ and the $\ell$ such that $\ell_{0}\equiv\ell\eta/v$
is a constant and $\ell\eta^{2}/v\to0$ then
\begin{align}
\frac{\partial\Gamma^{\left(2\right)}/VT}{\partial v} & =\ell_{0}m^{3},\\
\frac{\delta\Gamma^{\left(2\right)}}{\delta\Delta_{G}\left(z,w\right)} & =0,
\end{align}
in addition to the Ward identity and the $\Delta_{H}$ equation of motion. In practice, in the symmetry broken
phase, one simply discards the first equation of motion and solves
the second one in conjunction with the Ward identity, which suffices
to give a closed system. In the symmetric phase $v=0$ and the Ward
identity is trivial, but $\Gamma^{\left(2\right)}$ also does not
depend linearly on $v$, hence one can take the previous equations
of motion with $\ell_{0}=0$. Note that we can, and do, keep a nonzero
$m_{G}^{2}$ in the intermediate stages of the computation to serve
as an infrared regulator.

To recap the procedure: first we define a symmetry improved effective
action using Lagrange multipliers and compute the equations of motion.
Second, note that the equations of motion are singular when the constraints
are applied. Third, regulate the singularity by slightly violating
the constraint. Fourth, pass to a suitable limit where violation of
the constraint tends to zero. We require the limiting procedure to
be universal in the sense that no additional data (arbitrary forms
of the Lagrange multiplier fields) need be introduced into the theory.

We now extend this logic to the 3PI case. To that end we introduce
the symmetry improved 3PIEA
\begin{align}
\tilde{\Gamma}^{\left(3\right)} & =\Gamma^{\left(3\right)}+\frac{i}{2}\int_{x}\ell_{A}^{d}\left(x\right)\mathcal{W}_{c}^{A\left(1\right)}\left(x\right)\left[P_{T}\left(\varphi,x\right)\right]_{cd}\nonumber \\
 & -Bf\left[\mathcal{W}_{cd}^{A\left(1\right)}\right],
\end{align}
where the second term is the same as the 2PI symmetry improvement
term and the third term contains the extended symmetry improvement.
$B$ is the new Lagrange multiplier and $f\left[\mathcal{W}_{cd}^{A\left(1\right)}\right]$
is an arbitrary functional which vanishes if and only if its argument
vanishes. Substituting the SSB ansatz we obtain
\begin{equation}
\tilde{\Gamma}^{\left(3\right)}=\Gamma^{\left(3\right)}-\ell\mathcal{W}_{1}-Bf\left[\mathcal{W}_{2}\right].
\end{equation}
The equations of motion are
\begin{align}
\frac{\partial\Gamma^{\left(3\right)}/VT}{\partial v} & =\ell_{0}m^{3}+B\int_{xz}\frac{\delta f}{\delta\mathcal{W}_{2}\left(x,y\right)}\bar{V}\left(x,y,z\right),\label{eq:si3pi-v-eom}\\
\frac{\delta\Gamma^{\left(3\right)}}{\delta\Delta_{G}\left(r,s\right)} & =-B\int_{xy}\frac{\delta f}{\delta\mathcal{W}_{2}\left(x,y\right)}\Delta_{G}^{-1}\left(x,r\right)\Delta_{G}^{-1}\left(s,y\right),\label{eq:si3pi-deltaG-eom}\\
\frac{\delta\Gamma^{\left(3\right)}}{\delta\Delta_{H}\left(r,s\right)} & =B\int_{xy}\frac{\delta f}{\delta\mathcal{W}_{2}\left(x,y\right)}\Delta_{H}^{-1}\left(x,r\right)\Delta_{H}^{-1}\left(s,y\right),\label{eq:si3pi-deltaN-eom}\\
\frac{\delta\Gamma^{\left(3\right)}}{\delta\bar{V}\left(r,s,t\right)} & =vB\int_{xyz}\frac{\delta f}{\delta\mathcal{W}_{2}\left(x,y\right)}\nonumber \\
 & \times\delta^{\left(4\right)}\left(x-r\right)\delta^{\left(4\right)}\left(y-s\right)\delta^{\left(4\right)}\left(z-t\right),\label{eq:si3pi-barV-eom}\\
\frac{\delta\Gamma^{\left(3\right)}}{\delta V_{N}\left(r,s,t\right)} & =0,\label{eq:si3pi-VN-eom}\\
\mathcal{W}_{1} & =0,\label{eq:si3pi-deltaG-WI}\\
\mathcal{W}_{2} & =0,\label{eq:si3pi-barV-WI}
\end{align}
where we already take the previous limiting procedure to eliminate
$\ell$ and $\mathcal{W}_{1}$. In \eqref{eq:si3pi-barV-eom} we have
inserted a factor of $1=\int_{z}\delta^{\left(4\right)}\left(z-t\right)$
for later convenience. Now we devise a limiting procedure such that
the right hand sides of two of \eqref{eq:si3pi-deltaG-eom}, \eqref{eq:si3pi-deltaN-eom}
and \eqref{eq:si3pi-barV-eom} vanish. The remaining equation must
be chosen so that it can be replaced by the constraint \eqref{eq:si3pi-barV-WI}
and still give a closed system. Note that \eqref{eq:si3pi-barV-eom}
cannot be eliminated because there is not enough information to reconstruct
$\bar{V}$ from $\Delta_{G/H}$ using $\mathcal{W}_{2}$. Thus we
must eliminate either $\Delta_{G}$ or $\Delta_{H}$, or else artificially
restrict the form of $\bar{V}$.

We show that the desired simplification of the equations of motion
can be achieved without restricting $\bar{V}$ under the assumption
that $\delta f/\delta\mathcal{W}_{2}\left(x,y\right)$ is a spacetime
independent constant. Note that this is not required by Poincar\'e invariance
(only the weaker condition $\delta f/\delta\mathcal{W}_{2}\left(x,y\right)=g\left(\left|x-y\right|^{2}\right)$
is mandated). We temporarily adopt this assumption without further
explanation, though in Appendix \ref{sub:The-d'Alembert-formalism}
we will show that it can be justified by the introduction of the \emph{d'Alembert
formalism}.

Computing the left hand side of \eqref{eq:si3pi-barV-eom} using \eqref{eq:3piea}
and displaying only the two loop terms explicitly we obtain
\begin{align}
\frac{\delta\Gamma^{\left(3\right)}}{\delta\bar{V}\left(r,s,t\right)} & =-\frac{\hbar^{2}}{2}\left(N-1\right)\nonumber \\
 & \times\int_{xyz}\left(\bar{V}\left(x,y,z\right)+\frac{\lambda v}{3}\delta\left(x-y\right)\delta\left(x-z\right)\right)\nonumber \\
 & \times\Delta_{H}\left(x,r\right)\Delta_{G}\left(y,s\right)\Delta_{G}\left(z,t\right)+\mathcal{O}\left(\hbar^{3}\right).
\end{align}
Without symmetry improvement one sets this quantity to zero, giving
an equation equivalent to the one we derived in the previous section,
\eqref{eq:3pi-vertex-eom} (up to a group theory factor, since the
variables in the one case are $\Delta_{ab}$ and $V$
and in the other $\Delta_{G}$, $\Delta_{H}$ and $\bar{V}$). This
equation is now modified by the symmetry improvement to
\begin{multline}
-\frac{\hbar^{2}}{2}\left(N-1\right)\int_{xyz}\left(\bar{V}\left(x,y,z\right)+\frac{\lambda v}{3}\delta\left(x-y\right)\delta\left(x-z\right)\right)\\
\times\Delta_{H}\left(x,r\right)\Delta_{G}\left(y,s\right)\Delta_{G}\left(z,t\right)+\mathcal{O}\left(\hbar^{3}\right)\\
=vB\int_{xyz}\frac{\delta f}{\delta\mathcal{W}_{2}\left(x,y\right)}\delta^{\left(4\right)}\left(x-r\right)\delta^{\left(4\right)}\left(y-s\right)\delta^{\left(4\right)}\left(z-t\right).
\end{multline}
Convolving with the inverse propagators $\Delta_{H}^{-1}\Delta_{G}^{-1}\Delta_{G}^{-1}$
gives
\begin{multline}
-\frac{\hbar^{2}}{2}\left(N-1\right)\left(\bar{V}\left(a,b,c\right)+\frac{\lambda v}{3}\delta\left(a-b\right)\delta\left(a-c\right)\right)+\mathcal{O}\left(\hbar^{3}\right)\\
=vB\int_{xyz}\frac{\delta f}{\delta\mathcal{W}_{2}\left(x,y\right)}\Delta_{H}^{-1}\left(x,a\right)\Delta_{G}^{-1}\left(y,b\right)\Delta_{G}^{-1}\left(z,c\right)\\
=\left[B\int_{xy}\frac{\delta f}{\delta\mathcal{W}_{2}\left(x,y\right)}\Delta_{H}^{-1}\left(x,a\right)\Delta_{G}^{-1}\left(y,b\right)\right]\mathcal{W}_{1}.
\end{multline}
The right hand side now vanishes due to \eqref{eq:si3pi-deltaG-WI}.
With the regulator \eqref{eq:regulated-2pi-WI} in place, we have
that the symmetry improvement term in the $\bar{V}$ equation of motion
vanishes faster than the naive $B\delta f/\delta\mathcal{W}_{2}$ scaling
manifest in \eqref{eq:si3pi-barV-eom}. Schematically, the right hand
side scales as $B\left(\delta f/\delta\mathcal{W}_{2}\right)m_{H}^{2}m_{G}^{4}v\sim B\left(\delta f/\delta\mathcal{W}_{2}\right)m_{H}^{8}\eta^{2}/v$.
So long as $B\delta f/\delta\mathcal{W}_{2}$ does not blow up as fast
as $\eta^{-2}$ as $\eta\to0$ the symmetry improvement has no effect
on $\bar{V}$.

\begin{figure}
\includegraphics[width=0.5\columnwidth]{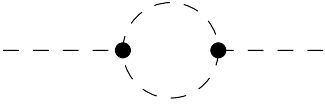}

\protect\caption{\label{fig:sunset-graph}Sunset self-energy graph}
\end{figure}

Now we investigate the Goldstone propagator. Substituting \eqref{eq:3piea}
into \eqref{eq:si3pi-deltaG-eom} we find the symmetry improved equation
of motion for $\Delta_{G}$:

\begin{equation}
\Delta_{G}^{-1}\left(r,s\right)=\Delta_{0G}^{-1}\left(r,s\right)-\tilde{\Sigma}_{G}\left(r,s\right),
\end{equation}
where we have defined the 3PI symmetry improved self-energy
\begin{align}
\tilde{\Sigma}_{G}\left(r,s\right) & \equiv\frac{2i}{\hbar\left(N-1\right)}\left[\frac{\delta\Gamma_{3}}{\delta\Delta_{G}\left(r,s\right)}\right.\nonumber \\
 & \left.+B\int_{xy}\frac{\delta f}{\delta\mathcal{W}_{2}\left(x,y\right)}\Delta_{G}^{-1}\left(x,r\right)\Delta_{G}^{-1}\left(s,y\right)\right].
\end{align}
Substituting in $\Gamma_{3}^{\left(3\right)}$ to two loop order we
find
\begin{align}
\tilde{\Sigma}_{G}\left(r,s\right) & =\frac{i\hbar\lambda}{6}\mathrm{Tr}\left[\left(N+1\right)\Delta_{G}+\Delta_{H}\right]\delta^{\left(4\right)}\left(r-s\right)\nonumber \\
 & -i\hbar\int_{abcd}\left(\bar{V}\left(a,b,r\right)+\frac{2\lambda v}{3}\delta^{\left(4\right)}\left(a-r\right)\delta^{\left(4\right)}\left(b-r\right)\right)\nonumber \\
 & \times\bar{V}\left(c,d,s\right)\Delta_{H}\left(a,c\right)\Delta_{G}\left(b,d\right)\nonumber \\
 & +\frac{2i}{\hbar\left(N-1\right)}B\int_{xy}\frac{\delta f}{\delta\mathcal{W}_{2}\left(x,y\right)}\Delta_{G}^{-1}\left(x,r\right)\Delta_{G}^{-1}\left(s,y\right)\nonumber \\
 & +\mathcal{O}\left(\hbar^{2}\right).
\end{align}
The first term corresponds to the Hartree-Fock diagram (Figure \ref{fig:hartree-superdaisies},
far left); the second term corresponds to the sunset diagrams (Figure
\ref{fig:sunset-graph}) and the third term is the symmetry improvement
term. The equation of motion for $\Delta_{H}$ can be written in the
same form with a suitable definition for a symmetry improved self
energy $\tilde{\Sigma}_{H}\left(r,s\right)$, where the symmetry improvement
term now has the form $\sim B\int\left(\delta f/\delta\mathcal{W}_{2}\right)\Delta_{H}^{-1}\Delta_{H}^{-1}$.

If we assume $\delta f/\delta\mathcal{W}_{2}$ is constant, we find that
$\tilde{\Sigma}_{G}$ and $\tilde{\Sigma}_{H}$ scale as $\left(B\delta f/\delta\mathcal{W}_{2}\right)m_{G}^{4}\sim\left(B\delta f/\delta\mathcal{W}_{2}\right)m_{H}^{6}\eta^{2}/v^{2}$
and $\left(B\delta f/\delta\mathcal{W}_{2}\right)m_{H}^{4}\sim\left(B\delta f/\delta\mathcal{W}_{2}\right)m_{H}^{4}\eta^{0}$
respectively. Thus, by choosing a regulator such that $B\delta f/\delta\mathcal{W}_{2}$
goes to a finite limit, the equations of motion for $\bar{V}$ and
$\Delta_{G}$ are unmodified and the equation of motion for $\Delta_{H}$
is modified by a finite term. This is the desired limiting procedure.
Adopting it gives the final set of equations of motion:
\begin{align}
\frac{\delta\Gamma^{\left(3\right)}}{\delta\Delta_{G}\left(r,s\right)} & =0,\label{eq:si3pi-DeltaG-eom}\\
\frac{\delta\Gamma^{\left(3\right)}}{\delta\bar{V}\left(r,s,t\right)} & =0,\label{eq:si3pi-Vbar-eom}\\
\frac{\delta\Gamma^{\left(3\right)}}{\delta V_{N}\left(r,s,t\right)} & =0,\label{eq:si3pi-VN-eom-1}\\
\mathcal{W}_{1} & =0,\label{eq:si3pi-w1-eom}\\
\mathcal{W}_{2} & =0.\label{eq:si3pi-w2-eom}
\end{align}

\section{Renormalization\label{sec:Renormalization}}

Here we undertake a general description of the renormalization problem
at zero temperature. Our detailed considerations follow in Sections
\ref{sub:Two-loop-truncation} and \ref{sub:Three-loop-truncation}
for two and three loop truncations respectively. Finite temperature
results are given for the Hartree-Fock approximation in Section \ref{sec:Hartree-approximation}.
The two loop renormalization of the theory in Section \ref{sub:Two-loop-truncation}
is non-trivial already even though the vertex equation of motion can
be solved trivially. This is because the symmetry improvement breaks
the $n$PIEA equivalence hierarchy by modifying the Higgs equation
of motion.

Generically, modifications of the equations of motion following from
the 2PIEA will lead to an inconsistency of the renormalization procedure
since the 2PIEA is self-consistently complete at two loop order (in
the action, i.e. one loop order in the equations of motion). However,
we will see that the wavefunction and propagator renormalization constants
(normally trivial in $\phi^{4}$ at one loop) provide the extra freedom
required to obtain consistency. Then in Section \ref{sub:Three-loop-truncation}
we will renormalize the theory at three loops. Non-perturbative counter-term
calculations are generally much more difficult than the analogous
perturbative calculations, hence many of the manipulations were performed
in a supplemental \noun{Mathematica} notebook \citep{supp}. The results
of this section are finite equations of motion for renormalized quantities
which must be solved numerically. We leave the numerical implementation
to future work, except in the case of the Hartree-Fock approximation.

We wish to demonstrate the renormalizability of the equations of motion
\eqref{eq:si3pi-DeltaG-eom}-\eqref{eq:si3pi-w2-eom}. First we examine
the symmetric phase, since on physical grounds SSB is irrelevant to
renormalizability. In the symmetric phase $v=0$ and the Ward identity
\eqref{eq:si3pi-w1-eom} is trivially satisfied, while \eqref{eq:si3pi-w2-eom}
requires $\Delta_{G}=\Delta_{H}$ as expected. Further, iteration
shows that \eqref{eq:si3pi-Vbar-eom} and \eqref{eq:si3pi-VN-eom-1}
have the solution $\bar{V}=V_{N}=0$ as expected on general grounds:
there is no three point vertex in the symmetric phase. As a result,
the symmetry improved 3PIEA in the symmetric phase is equivalent to
the ordinary 2PIEA
\begin{equation}
\tilde{\Gamma}^{\left(3\right)}\left[\varphi=0,\Delta,V=0\right]=\Gamma^{\left(2\right)}\left[\varphi=0,\Delta\right],
\end{equation}
which is known to be renormalizable, either by an implicit construction
involving Bethe-Salpeter integral equations or an explicit algebraic
BPHZ (Bogoliubov-Parasiuk-Hepp-Zimmerman) style construction which
has nontrivial consistency requirements (but which has been shown
to be equivalent to the Bethe-Salpeter method) \citep{Berges2005,Fejos2008,*Patkos2009,Marko2012}.
Thus, only divergences arising from nonzero $V$
pose any new conceptual problems.

We will extend the BPHZ style procedure of \citep{Fejos2008,Patkos2008,*Patkos2009},
which was adapted to symmetry improved 2PIEA by \citep{Pilaftsis2013}
and to 3PIEA for three dimensional pure glue QCD (without symmetry
improvement) by \citep{York2012}. The essence of the procedure is
quite simple. Consider for example the quadratically divergent integral
$\int_{q}i\Delta_{G/H}\left(q\right)$. Since $\Delta_{G/H}\left(q\right)$
is determined self-consistently this is a complicated integral which
must be evaluated numerically. However, the UV behavior of the propagator
should approach $q^{-2}$ as $q\to\infty$. (Note that Weinberg's
theorem \citep{Weinberg1960} implies that the self-consistent propagators
have this form up to powers of logarithms \citep{Patkos2008}, though
renormalization group theory shows the true large-momentum behavior
of the propagators is a power law with an anomalous dimension. This
implies that a truncated $n$PIEA does not effect a resummation of large
logarithms.) Now we can add and subtract an integral with the same
UV asymptotics:
\begin{align}
\int_{q}i\Delta_{G/H}\left(q\right) & =\int_{q}\left[i\Delta_{G/H}\left(q\right)-\frac{i}{q^{2}-\mu^{2}+i\epsilon}\right]\nonumber \\
 & +\int_{q}\frac{i}{q^{2}-\mu^{2}+i\epsilon},
\end{align}
where $\mu$ is an arbitrary mass subtraction scale (not a cutoff
scale). The first term is now only logarithmically divergent and the
second term can be evaluated analytically in a chosen regularization
scheme such as dimensional regularization. A further subtraction of
this kind can render the first term finite.

We write the renormalized propagators as
\begin{align}
\Delta_{G/H}^{-1} & =p^{2}-m_{G/H}^{2}-\Sigma_{G/H}\left(p\right),\\
\Sigma_{G/H}\left(p\right) & =\Sigma_{G/H}^{a}\left(p\right)+\Sigma_{G/H}^{0}\left(p\right)+\Sigma_{G/H}^{r}\left(p\right),
\end{align}
where $m_{G/H}$ is the physical mass and the (renormalized) self-energies
have been separated into pieces according to their asymptotic behavior:
$\Sigma_{G/H}^{a}\left(p\right)\sim p^{2}\left(\ln p\right)^{c_{1}}$,
$\Sigma_{G/H}^{0}\sim\left(\ln p\right)^{c_{2}}$ and $\Sigma_{G/H}^{r}\sim p^{-2}$
as $p\to\infty$ respectively. The pole condition requires $\Sigma_{G/H}\left(p^{2}=m_{G/H}^{2}\right)=0$.
We also introduce the auxiliary propagator $\Delta^{\mu}_{G/H}=\left(p^{2}-\mu^{2}-\Sigma_{G/H}^{a}\left(p\right)\right)^{-1}$.
The propagator $\Delta_{G/H}$ can be expanded in $\Delta^{\mu}_{G/H}$:
\begin{align}
\Delta_{G/H}\left(p\right) & =\Delta^{\mu}_{G/H}\left(p\right)\nonumber \\
 & +\left[\Delta^{\mu}_{G/H}\left(p\right)\right]^{2}\left(m_{G/H}^{2}-\mu^{2}+\Sigma_{G/H}^{0}+\Sigma_{G/H}^{r}\right)\nonumber \\
 & +\mathcal{O}\left(\left[\Delta^{\mu}_{G/H}\left(p\right)\right]^{3}\left[\Sigma_{G/H}^{0}\left(p\right)\right]^{2}\right).\label{eq:aux-propagator}
\end{align}
This allows us to extract the leading order asymptotics of diagrams
as $p\to\infty$.

We now do a similar analysis to isolate the leading asymptotics for
$V$ at large momentum. Suppressing $\mathrm{O}\left(N\right)$
indices we can write $V\left(p_{1},p_{2},p_{3}\right)=\lambda vf\left(\frac{p_{1}}{v},\frac{p_{2}}{v},\frac{p_{3}}{v}\right)$
where $p_{1}+p_{2}+p_{3}=0$. Now $V\to0$ as $v\to0$
implies that $f\left(\chi_{1},\chi_{2},\chi_{3}\right)\sim\chi^{\alpha}\left(\ln\chi\right)^{c_{3}}$,
where $\alpha<1$ and $\chi$ is representative of the largest scale
among $\chi_{1}$, $\chi_{2}$ and $\chi_{3}$. Now consider the vertex
equation of motion \eqref{eq:3pi-vertex-eom} or Figure \ref{fig:vertex-eom}.
The triangle graph goes like $\int_{\ell}\ell^{3\alpha-6}\left(\ln\ell\right)^{3c_{3}-3c_{1}}\sim\chi^{3\alpha-2}\left(\ln\chi\right)^{3c_{3}-3c_{1}}$
if $\alpha\neq2/3$ or $\left(\ln\chi\right)^{1+3c_{3}-3c_{1}}$ if
$\alpha=2/3$, which is dominated by the bubble graph which goes like
$\int_{\ell}\ell^{\alpha-4}\left(\ln\ell\right)^{c_{3}-2c_{1}}\sim\chi^{\alpha}\left(\ln\chi\right)^{c_{3}-2c_{1}}$
if $\alpha\neq0$ or $\left(\ln\chi\right)^{1+c_{3}-2c_{1}}$ if $\alpha=0$.
Thus to a leading approximation the large momentum behavior is obtained
by dropping the triangle graph from the equation of motion. This can
also be seen by taking $v\to0$ at fixed $p_{i}$ which suppresses
the triangle graph relative to the bubble graph.

\begin{figure}
\includegraphics[width=1\columnwidth]{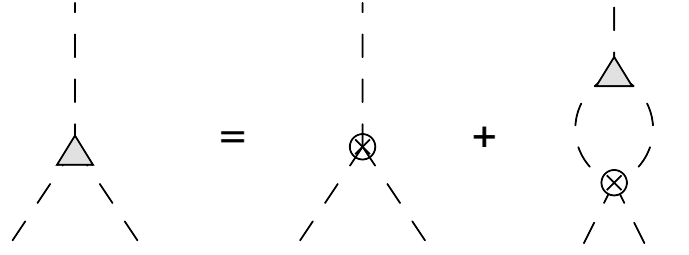}

\protect\caption{\label{fig:aux-vertex-eq}Defining equation for the auxiliary vertex
functions $\bar{V}^{\mu}$ and $V^{\mu}_{N}$, obtained by taking the
corresponding equations of motion for $\bar{V}$ and $V_{N}$, dropping
the triangle graphs, and making the replacements $\bar{V}\to\bar{V}^{\mu}$,
$V_{N}\to V^{\mu}_{N}$, and $\Delta_{G/H}\to\Delta^{\mu}_{G/H}$. The
filled triangle represents the auxiliary vertices and the lines represent
the auxiliary propagators. Crossed circles are bare vertices as before.}
\end{figure}

We now define auxiliary vertex functions $\bar{V}^{\mu}$ and $V^{\mu}_{N}$
which have the same asymptotic behavior as $\bar{V}$ and $V_{N}$
respectively, though depend only on the auxiliary propagators. We
define $\bar{V}^{\mu}$ and $V^{\mu}_{N}$ by taking the equations of
motion for $\bar{V}$ and $V_{N}$, dropping the triangle graphs,
and making the replacements $\bar{V}\to\bar{V}^{\mu}$, $V_{N}\to V^{\mu}_{N}$,
and $\Delta_{G/H}\to\Delta^{\mu}_{G/H}$. These equations are shown
in Figure \ref{fig:aux-vertex-eq}. This gives a pair of coupled linear
integral equations, analogous to the Bethe-Salpeter equations, for
$\bar{V}^{\mu}$ and $V^{\mu}_{N}$ which can be solved explicitly by
iteration. (Details of this calculation are presented in Section \ref{sub:Three-loop-truncation}).
Unfortunately the result is only analytically tractable in fewer than
1+3 dimensions, so we confine the analytical results depending on
the explicit forms of $\bar{V}^{\mu}$ and $V^{\mu}_{N}$ to this case.
In the physically most interesting case of 1+3 dimensions, $\bar{V}^{\mu}$
and $V^{\mu}_{N}$ must be numerically determined at the same time as
$\bar{V}$ and $V_{N}$.

By using these auxiliary propagators and vertices we can isolate the
divergent contributions to the equations of motion and so obtain the
required set of counter-terms to remove them.

\subsection{Two loop truncation\label{sub:Two-loop-truncation}}

The theory simplifies dramatically at two loop order. It follows from
\eqref{eq:3pi-vertex-eom} that at this order $V\to V_{0}$
(up to a renormalization). Substituting this into the action gives,
apart from the symmetry improvement terms, the standard 2PIEA. This
is an example of the equivalence hierarchy previously discussed. Another
simplification is that the logarithmic enhancement of the propagators
in the UV due to $\Sigma_{G/H}^{a}$ vanishes at this level ($\Sigma_{G/H}^{a}$
is generated by the diagram $\Phi_{5}$ appearing at three loop order).
In this case $\Delta_{G}^{\mu}=\Delta_{H}^{\mu}\equiv\Delta^{\mu}=\left(p^{2}-\mu^{2}\right)^{-1}$.
However, the reduction is not trivial because now the Higgs equation
of motion has been replaced by a Ward identity. The equations of motion
reduce to

\begin{align}
\Delta_{G}^{-1}\left(x,y\right) & =-\left(\partial_{\mu}\partial^{\mu}+m^{2}+\frac{\lambda}{6}v^{2}\right)\delta^{\left(4\right)}\left(x-y\right)\nonumber \\
 & -\frac{i\hbar}{6}\left(N+1\right)\lambda\Delta_{G}\left(x,x\right)\delta^{\left(4\right)}\left(x-y\right)\nonumber \\
 & -\frac{i\hbar}{6}\lambda\Delta_{H}\left(x,x\right)\delta^{\left(4\right)}\left(x-y\right)\nonumber \\
 & -\frac{i\hbar}{9}\lambda^{2}v^{2}\Delta_{H}\left(x,y\right)\Delta_{G}\left(x,y\right),\\
\Delta_{H}^{-1}\left(x,y\right) & =-\frac{\lambda v^{2}}{3}\delta^{\left(4\right)}\left(x-y\right)+\Delta_{G}^{-1}\left(x,y\right),\\
vm_{G}^{2} & =0.
\end{align}
The first line is the tree level term, the second and third lines
are the Hartree-Fock self-energies, the fourth line is the sunset
self-energy, and the last two lines are the Ward identities $\mathcal{W}_{2}$
and $\mathcal{W}_{1}$ respectively.

To renormalize the theory we regard all parameters heretofore as bare
parameters and introduce renormalized counterparts using the same
letters:
\begin{align}
\left(\phi,\varphi,v\right) & \to Z^{1/2}\left(\phi,\varphi,v\right),\label{eq:renorm-intro-wavefun-Z} \\
m^{2} & \to Z^{-1}Z_{\Delta}^{-1}\left(m^{2}+\delta m^{2}\right),\label{eq:renorm-intro-delta-m}\\
\lambda & \to Z^{-2}\left(\lambda+\delta\lambda\right),\label{eq:renorm-intro-delta-lambda}\\
\Delta & \to ZZ_{\Delta}\Delta,\label{eq:renorm-intro-Z_Delta}\\
V & \to Z^{-3/2}Z_{V}V\label{eq:renorm-intro-Z_V}.
\end{align}
Hereafter whenever we refer to a bare parameter we indicate this with
a subscript ``B,'' e.g. $m_{B}^{2}$ etc. The wavefunction renormalizations
for $\Delta$ and $V$ can be obtained from their definitions $\Delta\sim\left\langle \phi\phi\right\rangle $
and $\Delta\Delta\Delta V\sim\left\langle \phi\phi\phi\right\rangle $
respectively. Due to the presence of composite operators in the effective
action, additional counter-terms are required compared to the standard
perturbation theory: $\delta m_{0}^{2}$ and $\delta\lambda_{0}$
for terms in the bare action, $\delta m_{1}^{2}$ for one loop terms,
$\delta\lambda_{1}^{A}$ for terms of the form $\phi_{i}\phi_{i}\Delta_{jj}$,
$\delta\lambda_{1}^{B}$ for $\phi_{i}\phi_{j}\Delta_{ij}$ terms,
$\delta\lambda_{2}^{A}$ for $\Delta_{ii}\Delta_{jj}$ and $\delta\lambda_{2}^{B}$
for $\Delta_{ij}\Delta_{ij}$. Similarly, $\Delta$ and $V$
are given independent renormalization constants $Z_{\Delta}$ and
$Z_{V}$ respectively. We give terms in the sunset graphs a universal
$\delta\lambda$ counter-term.

The renormalized equations of motion are (see Appendix \ref{sec:Deriving-counter-terms-for-2-loops}
for more detail)
\begin{align}
\Delta_{G}^{-1}\left(p\right) & =ZZ_{\Delta}p^{2}-m^{2}-\delta m_{1}^{2}-Z_{\Delta}\frac{\lambda+\delta\lambda_{1}^{A}}{6}v^{2}\nonumber \\
 & -\frac{\hbar}{6}\left[\left(N+1\right)\lambda+\left(N-1\right)\delta\lambda_{2}^{A}+2\delta\lambda_{2}^{B}\right]Z_{\Delta}^{2}\mathcal{T}_{G}\nonumber \\
 & -\frac{\hbar}{6}\left(\lambda+\delta\lambda_{2}^{A}\right)Z_{\Delta}^{2}\mathcal{T}_{H}\nonumber \\
 & +i\hbar\left[\frac{\left(\lambda+\delta\lambda\right)v}{3}\right]^{2}Z_{\Delta}^{3}\mathcal{I}_{HG}\left(p\right),\label{eq:two-loop-Delta_G-eom-with-cts}\\
\Delta_{H}^{-1}\left(p\right) & =-Z_{\Delta}\frac{\left(\lambda+\delta\lambda\right)v^{2}}{3}+\Delta_{G}^{-1}\left(p\right),\label{eq:two-loop-Delta_N-eom-with-cts}\\
vm_{G}^{2} & =0,\label{eq:two-loop-Goldstone-thm-with-cts}
\end{align}
where for convenience we have defined
\begin{align}
\mathcal{T}_{G/H} & =\int_{q}i\Delta_{G/H}\left(q\right),\label{eq:tadpole-integral}\\
\mathcal{I}_{HG}\left(p\right) & =\int_{q}i\Delta_{H}\left(q\right)i\Delta_{G}\left(p-q\right).\label{eq:Ing-integral}
\end{align}
The tadpole integrals $\mathcal{T}_{G/H}$ correspond to the Hartree-Fock
graphs and $\mathcal{I}_{HG}\left(p\right)$ corresponds to the sunset
self-energy graph. From these we identify the self-energy parts

\begin{align}
\Sigma_{G/H}^{a}\left(p\right) & =\left(ZZ_{\Delta}-1\right)p^{2}=0,\\
\Sigma_{G/H}^{0}\left(p\right)+\Sigma_{G/H}^{r}\left(p\right) & =-i\hbar\left[\frac{\left(\lambda+\delta\lambda\right)v}{3}\right]^{2}Z_{\Delta}^{3}\mathcal{I}_{HG}\left(p\right).
\end{align}
Note that the Goldstone and Higgs self-energies are equal to this
order as a consequence of the vertex Ward identity. This is essentially
where our treatment differs from \citep{Pilaftsis2013}.

By using the auxiliary propagators to extract the divergences in $\mathcal{T}_{G/H}$
and $\mathcal{I}_{HG}\left(p\right)$ and absorbing them into the
counter-terms (see Appendix \ref{sec:Deriving-counter-terms-for-2-loops})
we find the finite equations of motion

\begin{align}
\Delta_{G}^{-1}\left(p\right) & =p^{2}-m^{2}-\frac{\lambda}{6}v^{2}-\frac{\hbar}{6}\left(N+1\right)\lambda\mathcal{T}_{G}^{\text{fin}}-\frac{\hbar}{6}\lambda\mathcal{T}_{H}^{\text{fin}}\nonumber \\
 & +i\hbar\left(\frac{\lambda v}{3}\right)^{2}\left[\mathcal{I}_{HG}^{\text{fin}}\left(p\right)-\mathcal{I}_{HG}^{\text{fin}}\left(m_{G}\right)\right],\label{eq:Goldstone-eom-finite}\\
\Delta_{H}^{-1}\left(p\right) & =p^{2}-m^{2}-\frac{\lambda v^{2}}{3}-\frac{\lambda}{6}v^{2}-\frac{\hbar}{6}\left(N+1\right)\lambda\mathcal{T}_{G}^{\text{fin}}\nonumber \\
 & -\frac{\hbar}{6}\lambda\mathcal{T}_{H}^{\text{fin}} +i\hbar\left(\frac{\lambda v}{3}\right)^{2}\left[\mathcal{I}_{HG}^{\text{fin}}\left(p\right)-\mathcal{I}_{HG}^{\text{fin}}\left(m_{H}\right)\right].\label{eq:Higgs-eom-finite}
\end{align}

The finite parts $\mathcal{T}_{G/H}^{\text{fin}}$ and $\mathcal{I}_{HG}^{\text{fin}}\left(p\right)$
are
\begin{align}
\mathcal{I}_{HG}^{\text{fin}}\left(p\right)	&=\mathcal{I}_{HG}\left(p\right)-\mathcal{I}^{\mu},\\
\mathcal{T}_{G/H}^{\text{fin}} &=\mathcal{T}_{G/H}-\mathcal{T}^{\mu}+i\left(m_{G/H}^{2}-\mu^{2}\right)\mathcal{I}^{\mu}\nonumber\\
 &-\int_{q}i\left[\Delta^{\mu}\left(q\right)\right]^{2}\Sigma^{\mu}\left(q\right),
\end{align}
where the auxiliary quantities are
\begin{align}
\mathcal{T}^{\mu} &=\int_{q}i\Delta^{\mu}\left(q\right),\\
\mathcal{I}^{\mu} &=\int_{q}\left[i\Delta^{\mu}\left(q\right)\right]^{2},\\
\Sigma^{\mu}\left(q\right) &=-i\hbar\left(\frac{\lambda v}{3}\right)^{2} \left[\int_{\ell}i\Delta^{\mu}\left(\ell\right)i\Delta^{\mu}\left(q+\ell\right)-\mathcal{I}^{\mu}\right].
\end{align}
(For details see Appendix \ref{sec:Deriving-counter-terms-for-2-loops}.)
These equations are the main result of this section. We expect they
could be solved numerically using techniques similar to \citep{Marko2012},
though we leave the numerical implementation for later work.

\subsection{Three loop truncation\label{sub:Three-loop-truncation}}

We consider now the three loop truncation of the effective action.
The vertex equation of motion is shown in Figure \ref{fig:vertex-eom},
and we have already argued that the leading asymptotics at large momentum
are captured by the auxiliary vertex defined by its equation of motion
in Figure \ref{fig:aux-vertex-eq}. Subtracting these two equations
we find that the right hand side is finite (indeed the auxiliary vertices
were constructed to guarantee this). Thus the problem of renormalizing
the vertex equation of motion reduces to the problem of renormalizing
the auxiliary vertex equation of motion.

\begin{figure}
\includegraphics[width=1\columnwidth]{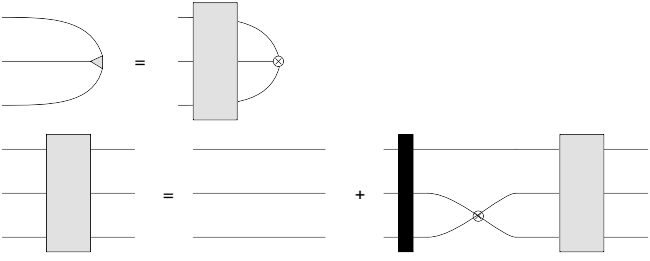}

\protect\caption{\label{fig:aux-vertex-soln-and-6pt-kernel}Solution for the auxiliary
vertex function in terms of a six point kernel $\mathcal{K}_{abcdef}$
which is represented by the blue box (the indices run from top to
bottom down the left side, then the right). The vertical black bar
in the kernel equation of motion represents symmetrization of the
external lines (with a factor of $1/3!$).}
\end{figure}

It is temporarily more convenient to go back to the $\mathrm{O}\left(N\right)$
covariant form we had before introducing the SSB ansatz. Introduce
the covariant auxiliary vertex $V^{\mu}_{abc}$ which is related to
$\bar{V}^{\mu}$ and $V^{\mu}_{N}$ by an equation analogous to \eqref{eq:ssb-vertex-ansatz}.
Iterating the equation of motion we find the solution
\begin{equation}
V^{\mu}_{abc}=\mathcal{K}_{abcdef}V_{0def},\label{eq:aux-vertex-soln}
\end{equation}
where the six point kernel $\mathcal{K}_{abcdef}$ obeys the Bethe-Salpeter
like equation
\begin{align}
\mathcal{K}_{abcdef} & =\delta_{ad}\delta_{be}\delta_{cf}+\frac{1}{3!}\sum_{\pi}\left(-\frac{3i\hbar}{2}\right)\delta_{\pi\left(a\right)h}\nonumber \\
 & \times W_{\pi\left(b\right)\pi\left(c\right)kg}\Delta^{\mu}_{ki}\Delta^{\mu}_{gj}\mathcal{K}_{hijdef},\label{eq:K-6pt-kernel-eom}
\end{align}
where $\sum_{\pi}$ is a sum over permutations of the incoming legs.
This equation is shown in Figure \ref{fig:aux-vertex-soln-and-6pt-kernel}.
\eqref{eq:K-6pt-kernel-eom} can be written
in a form that makes explicit all divergences (see Appendix \ref{sec:Auxiliary-vertex-and-renorm})
and replaces the bare vertex $W$ by a four point kernel $\mathcal{K}_{abcd}^{\left(4\right)}\sim\lambda/\left(1+\lambda\mathcal{I}^{\mu}\right)$.

In fewer than four dimensions $\mathcal{K}_{abcd}^{\left(4\right)}$
is finite and every correction to the tree level value is asymptotically
sub-dominant. Thus the leading term at large momentum is the tree level term
and, instead of the full auxiliary vertex as we have defined it, one can
simply take $V^{\mu}_{abc}=V_{0abc}$, dramatically simplifying the
renormalization theory. A similar simplification happens to the auxiliary
propagator due to the logarithmic (rather than quadratic) divergence of
$\Phi_{5}$-generated self-energy in $<1+3$ dimensions. This confirms statements
made in the literature (supported by numerical evidence though without proof,
to our knowledge) to the effect that the asymptotic behavior of Green functions
is free (e.g. \citep{York2012}).

Unfortunately, the situation is much more difficult in four dimensions
and the renormalization of the $n$PIEA for $n\geq 3$ in $d>3$ remains an open
problem, both in general and in the present case. The problem can be seen
from the behaviour of the auxiliary vertex which is discussed further in Appendix \ref{sec:Auxiliary-vertex-and-renorm}.
For the sake of obtaining analytical results we restrict the rest of this
section to $<1+3$ dimensions. The renormalization of the $1+3$
dimensional case is left to future work.

We derive the counter-terms for $1+2$ dimensions in Appendix \ref{sec:Deriving-Counter-terms-for-3-loops}.
The are only two interesting comments about this derivation: the first
is that we require an additional (non-universal) counter-term for
the sunset graph linear in $V$; the second is that,
consistent with the super-renormalizability of $\phi^{4}$ theory
in $1+2$ dimensions, only $\delta m_{1}^{2}$ is required to UV-renormalize
the theory. Every other counter-term is finite and exists solely to
maintain the pole condition for the Higgs propagator despite
the vertex Ward identity. The resulting finite equations of motion
are

\begin{align}
\Delta_{G}^{-1} & =-\left(\partial_{\mu}\partial^{\mu}+m^{2}+\frac{\lambda}{6}v^{2}\right)\nonumber \\
 & -\left[\Sigma_{G}^{0}\left(p\right)-\Sigma_{G}^{0}\left(m_{G}\right)\right],\label{eq:three-loop-geom-finite}
\end{align}
for the Goldstone propagator,

\begin{align}
\bar{V} & =-\frac{\lambda v}{3}\nonumber \\
 & +i\hbar\left[V_{N}\left(\bar{V}\right)^{2}\left(\Delta_{H}\right)^{2}\Delta_{G}+\left(\bar{V}\right)^{3}\Delta_{H}\left(\Delta_{G}\right)^{2}\right]\nonumber \\
 & +\frac{i\hbar\lambda}{6}\left[V_{N}\left(\Delta_{H}\right)^{2}+\left(N+1\right)\bar{V}\left(\Delta_{G}\right)^{2}+4\bar{V}\Delta_{G}\Delta_{H}\right],\label{eq:three-loop-Vbar-eom-finite}
\end{align}
for the Higgs-Goldstone-Goldstone vertex, and

\begin{align}
V_{N} & =-\lambda v\nonumber \\
 & +i\hbar\left[\left(N-1\right)\left(\bar{V}\right)^{3}\left(\Delta_{G}\right)^{3}+\left(V_{N}\right)^{3}\left(\Delta_{H}\right)^{3}\right]\nonumber \\
 & +\frac{i\hbar\lambda}{2}\left[\left(N-1\right)\bar{V}\Delta_{G}\Delta_{G}+3V_{N}\left(\Delta_{H}\right)^{2}\right],\label{eq:three-loop-V_N-eom-finite}
\end{align}
for the triple Higgs vertex.

\begin{figure}
\includegraphics[width=1\columnwidth]{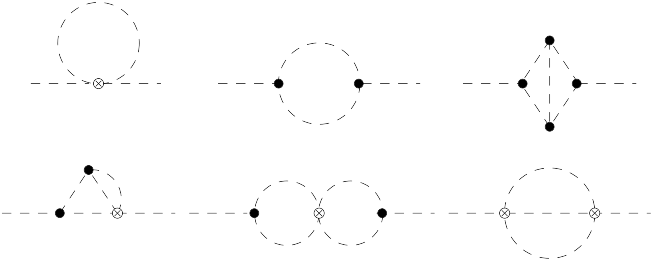}

\protect\caption{\label{fig:two-loop-self-energy-graphs}Feynman graph topologies appearing
in the self-energy function $\Sigma_{G}^{0}\left(p\right)$ in \eqref{eq:three-loop-goldstone-self-energy-finite}.}
\end{figure}

The finite Goldstone self-energy is
\begin{widetext}
\begin{align}
-\Sigma_{G}^{0}\left(p\right) & =-\frac{\hbar}{6}\left(N+1\right)\lambda\left(\mathcal{T}_{G}-\mathcal{T}^{\mu}\right)-\frac{\hbar}{6}\lambda\left(\mathcal{T}_{H}-\mathcal{T}^{\mu}\right)\nonumber \\
 & -i\hbar\left[-2\frac{\lambda v}{3}-\bar{V}\right]\Delta_{H}\Delta_{G}\bar{V}+\hbar^{2}\left[V_{N}\left(\bar{V}\right)^{3}\left(\Delta_{H}\right)^{3}\left(\Delta_{G}\right)^{2}+\left(\bar{V}\right)^{4}\Delta_{H}\Delta_{H}\left(\Delta_{G}\right)^{3}\right]\nonumber \\
 & +\frac{\hbar^{2}\lambda}{3}\left[\bar{V}V_{N}\left(\Delta_{H}\right)^{3}\Delta_{G}+\left(N+1\right)\bar{V}\bar{V}\Delta_{H}\left(\Delta_{G}\right)^{3}+3\bar{V}\bar{V}\left(\Delta_{G}\right)^{2}\Delta_{H}\Delta_{H}\right]\nonumber \\
 & +\frac{\hbar^{2}\lambda^{2}}{18}\left[\left(N+1\right)\left(\Delta_{G}\right)^{3}+\Delta_{H}\Delta_{H}\Delta_{G}-\left(N+2\right)\mathcal{B}^{\mu}\right],\label{eq:three-loop-goldstone-self-energy-finite}
\end{align}

\end{widetext}
where the \noun{bball} integral is $\mathcal{B}^{\mu}=\int_{qp}\Delta^{\mu}\left(q\right)\Delta^{\mu}\left(p\right)\Delta^{\mu}\left(p+q\right)$.
The graph topologies are shown in Figure \ref{fig:two-loop-self-energy-graphs}.
Finally, the Higgs equation of motion is
\begin{align}
\Delta_{H}^{-1}\left(p\right) & =\left(m_{G}^{2}+\Sigma_{G}^{0}\left(m_{H}\right)-m_{H}^{2}\right)\nonumber \\
 & \times\frac{\bar{V}\left(p,-p,0\right)}{\bar{V}\left(m_{H},-m_{H},0\right)}+\Delta_{G}^{-1}\left(p\right).\label{eq:three-loop-Higgs-eom-finite}
\end{align}
The unusual form of this equation is a result of the pole condition
$\Delta_{H}^{-1}\left(m_{H}\right)=0$. We defer the numerical implementation
of these equations to future work.

\section{Solution of the Hartree-Fock approximation\label{sec:Hartree-approximation}}

In the Hartree-Fock approximation one drops the $\mathcal{I}_{HG}\left(p\right)$
term in the two loop equations of motion, or equivalently drops the
sunset diagram. In this case the problem simplifies dramatically because
the self-energy is momentum independent. The machinery of the auxiliary
propagators introduced previously is now unnecessary and $\mathcal{T}_{G/H}=\mathcal{T}_{G/H}^{\infty}+\mathcal{T}_{G/H}^{\text{fin}}$
can be written as the sum of divergent and finite parts which can
be evaluated in closed form. In the Matsubara formalism at finite
temperature $T$ the time contour is taken on the imaginary axis with
periodic boundary conditions of period $-i\beta$, where $\beta=1/T$.
Integration over the timelike momentum component $p^{0}$ becomes
a sum over discrete Matsubara frequencies $\omega_{n}=2\pi n/\beta$,
$n=0,\pm1,\pm2,\cdots$. Using standard tricks \citep{Kapusta2006}
the sum over frequencies can be performed, giving
\begin{align}
\mathcal{T}_{G/H}^{\infty} & =-\frac{m_{G/H}^{2}}{16\pi^{2}}\left[\frac{1}{\epsilon}-\gamma+1+\ln\left(4\pi\right)\right]+\mathcal{O}\left(\epsilon\right),\label{eq:divergent-part-tadpole-msbar}\\
\mathcal{T}_{G/H}^{\text{fin}} & =\mathcal{T}_{G/H}^{\text{vac}}+\mathcal{T}_{G/H}^{\text{th}},\label{eq:finite-part-tadpole}\\
\mathcal{T}_{G/H}^{\text{vac}} & =\frac{m_{G/H}^{2}}{16\pi^{2}}\ln\left(\frac{m_{G/H}^{2}}{\mu^{2}}\right),\label{eq:finite-vacuum-part-tadpole}\\
\mathcal{T}_{G/H}^{\text{th}} & =\int_{\boldsymbol{q}}\frac{1}{\omega_{\boldsymbol{q}}}\frac{1}{\mathrm{e}^{\beta\omega_{\boldsymbol{q}}}-1},\label{eq:bose-einstein-integral}
\end{align}
where the divergent and finite vacuum parts have been evaluated using
$\overline{\text{MS}}$ in $d=4-2\epsilon$ dimensions at the renormalization
point $\mu$ (note that \citep{Pilaftsis2013} adopt a slightly different
convention for $\mathcal{T}_{G/H}^{\text{vac}}$ which amounts to
a redefinition of $\mu$ not affecting physical results). $\gamma\approx0.577$
is the Euler-Mascheroni constant. In the thermal part $\boldsymbol{q}$
is the spatial momentum vector and $\omega_{\boldsymbol{q}}=\sqrt{\boldsymbol{q}^{2}+m_{G/H}^{2}}$.
We substitute these expressions into the equations of motion \eqref{eq:two-loop-Delta_G-eom-with-cts}-\eqref{eq:two-loop-Goldstone-thm-with-cts}
and demand that the kinematically distinct divergences proportional
to $v$, $\mathcal{T}_{G/H}^{\text{vac}}$, $\mathcal{T}_{G/H}^{\text{th}}$
independently vanish. The other renormalization conditions are that
residue of the pole of the propagator equals one, which requires $ZZ_{\Delta}=1$,
and that the tree level relation $m_{H}^{2}=\frac{\lambda v^{2}}{3}+m_{G}^{2}$
holds at zero temperature. These conditions determine the renormalization
constants
\begin{align}
Z=Z_{\Delta} & =1,\\
\delta m_{1}^{2} & =\frac{\left(N+2\right)\hbar\lambda m^{2}}{96\pi^{2}\epsilon}\frac{\left(\epsilon\kappa+1\right)}{1-\frac{\hbar\lambda\left(N+2\right)\left(\epsilon\kappa+1\right)}{96\pi^{2}\epsilon}},\\
\delta\lambda_{1}^{A} & =\frac{\left(N+4\right)\lambda}{\left(N+2\right)m^{2}}\delta m_{1}^{2},\\
\delta\lambda_{2}^{A} & =\delta\lambda_{2}^{B}=\frac{N+2}{N+4}\delta\lambda_{1}^{A},
\end{align}
where $\kappa\equiv1-\gamma+\ln4\pi\approx2.95$. Note that the undetermined
constant $\delta\lambda$ can be consistently set to zero at this
order. The finite equations of motion are
\begin{align}
m_{G}^{2} & =m^{2}+\frac{\lambda}{6}v^{2}+\frac{\hbar\lambda}{6}\left(N+1\right)\mathcal{T}_{G}^{\text{fin}}+\frac{\hbar\lambda}{6}\mathcal{T}_{H}^{\text{fin}},\label{eq:mg2-hartree-gap-eq}\\
m_{H}^{2} & =\frac{\lambda v^{2}}{3}+m_{G}^{2},\label{eq:mn2-hartree-gap-eq}\\
vm_{G}^{2} & =0.\label{eq:goldstone-theorem-hartree-gap-eq}
\end{align}
Finally, if we demand the zero temperature tree level relation $v^{2}\left(T=0\right)\equiv\bar{v}^{2}=-6m^{2}/\lambda$
we must set the renormalization point $\mu^{2}=\bar{m}_{H}^{2}\equiv m_{H}^{2}\left(T=0\right)=\lambda\bar{v}^{2}/3$.

The analogue of the equations of motion \eqref{eq:mg2-hartree-gap-eq}-\eqref{eq:goldstone-theorem-hartree-gap-eq}
corresponding to previous work on the symmetry improved 2PIEA is (\citep{Pilaftsis2013,Mao2013}
generalized to arbitrary $N$)

\begin{align}
m_{G}^{2} & =m^{2}+\frac{\lambda}{6}v^{2}+\frac{\hbar\lambda}{6}\left(N+1\right)\mathcal{T}_{G}^{\mathrm{fin}}+\frac{\hbar\lambda}{6}\mathcal{T}_{H}^{\mathrm{fin}},\label{eq:si2pi-goldstone-gap-eq}\\
m_{H}^{2} & =m^{2}+\frac{\lambda}{2}v^{2}+\frac{\hbar\lambda}{6}\left(N-1\right)\mathcal{T}_{G}^{\mathrm{fin}}+\frac{\hbar\lambda}{2}\mathcal{T}_{H}^{\mathrm{fin}},\label{eq:si2pi-higgs-gap-eq}\\
vm_{G}^{2} & =0.\label{eq:si2pi-goldstone-theorem}
\end{align}
Note that only the Higgs equation of motion differs, as expected.
In the standard formalism without symmetry improvement one replaces
\eqref{eq:si2pi-goldstone-theorem} with
\begin{equation}
0=v\left(m^{2}+\frac{\lambda}{6}v^{2}+\frac{\hbar\lambda}{6}\left(N-1\right)\mathcal{T}_{G}^{\mathrm{fin}}+\frac{\hbar\lambda}{2}\mathcal{T}_{H}^{\mathrm{fin}}\right).\label{eq:nosi-vev-eq}
\end{equation}

These equations of motion, or \emph{gap equations}, possess a phase
transition and a critical point where $m_{H}^{2}=m_{G}^{2}=v^{2}=0$.
Using the result for massless particles $\mathcal{T}_{G/H}^{\text{th}}\left(m_{G/H}=0\right)=T^{2}/12$,
we find the same value of the critical temperature
\begin{equation}
T_{\star}=\sqrt{\frac{12\bar{v}^{2}}{\hbar\left(N+2\right)}},
\end{equation}
independent of the formalism used. However, the order of the phase
transition differs in the three cases. This stands in contrast to
the large-$N$ approximation, which correctly determines the order
of the phase transition but gives a critical temperature larger by
a factor of $\sqrt{3/2}+\mathcal{O}\left(N^{-1}\right)$ (see \citep{Petropoulos2004,Mao2013}).

\begin{figure}
\includegraphics[width=1\columnwidth]{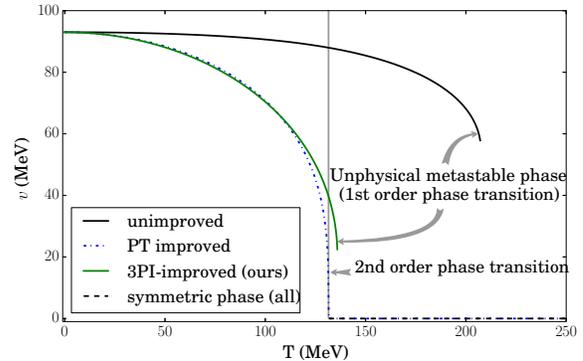}
\protect\caption{\label{fig:hartree-fock-vev-comparison}Expectation value of the scalar
field $v=\left\langle \phi\right\rangle $ as a function of temperature
$T$ computed in the Hartree-Fock approximation using the unimproved
2PIEA (solid black), the Pilaftsis and Teresi symmetry improved 2PIEA
(dash dotted blue) and our symmetry improved 3PIEA (solid green).
In the symmetric phase (dashed black) all methods agree. The vertical
grey line at $T\approx131.5\ \mathrm{MeV}$ corresponds to the critical
temperature which is the same in all methods.}
\end{figure}

\begin{figure}
\includegraphics[width=1\columnwidth]{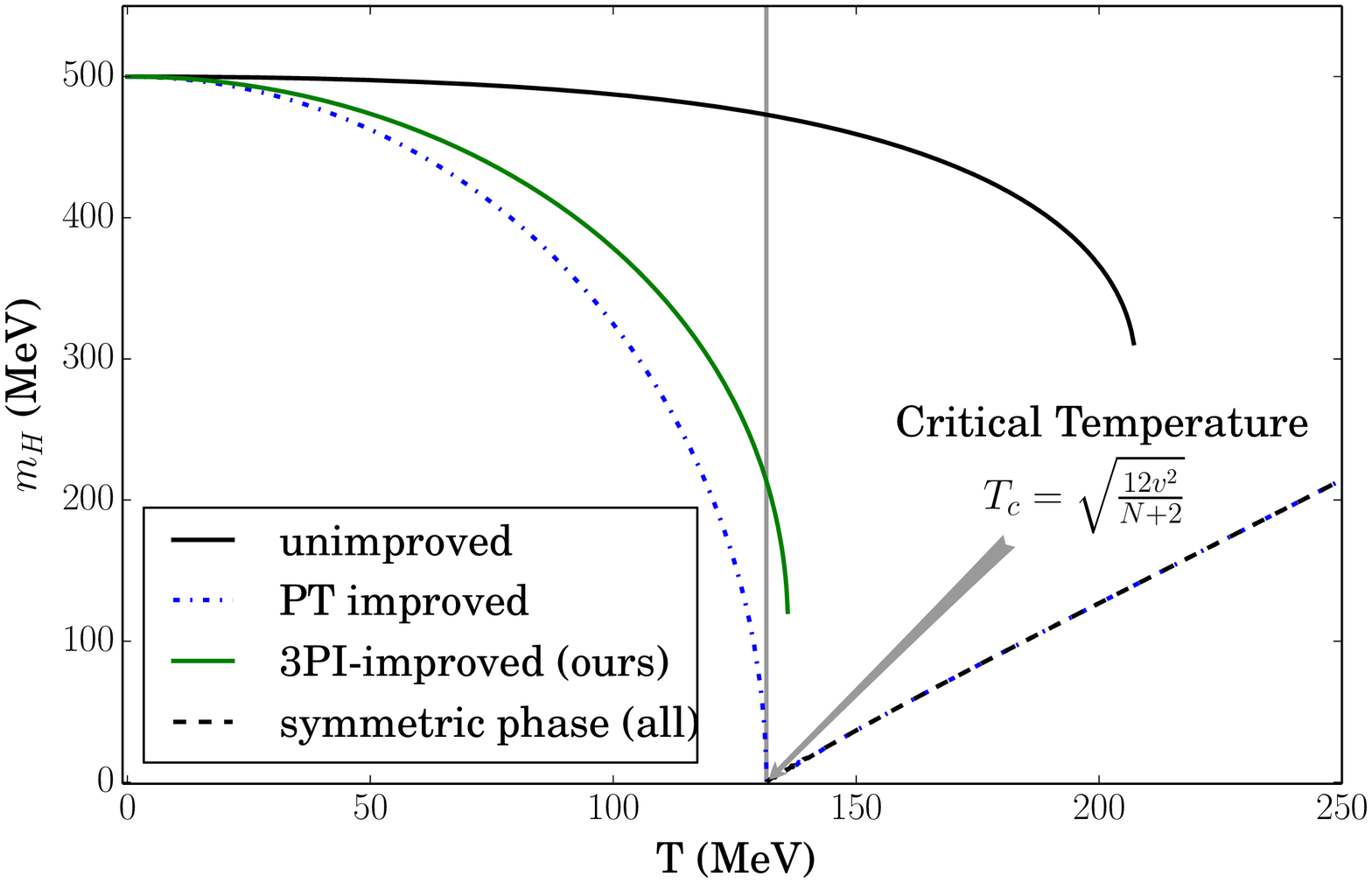}

\protect\caption{\label{fig:hartree-fock-mn-comparison}The Higgs mass $m_{H}$ as
a function of temperature $T$ computed in the Hartree-Fock approximation
using the unimproved 2PIEA (solid black), the Pilaftsis and Teresi
symmetry improved 2PIEA (dash dotted blue) and our symmetry improved
3PIEA (solid green). In the symmetric phase (dashed black) all methods
agree. The vertical grey line at $T\approx131.5\ \mathrm{MeV}$ corresponds
to the critical temperature which is the same in all methods.}
\end{figure}

\begin{figure}
\includegraphics[width=1\columnwidth]{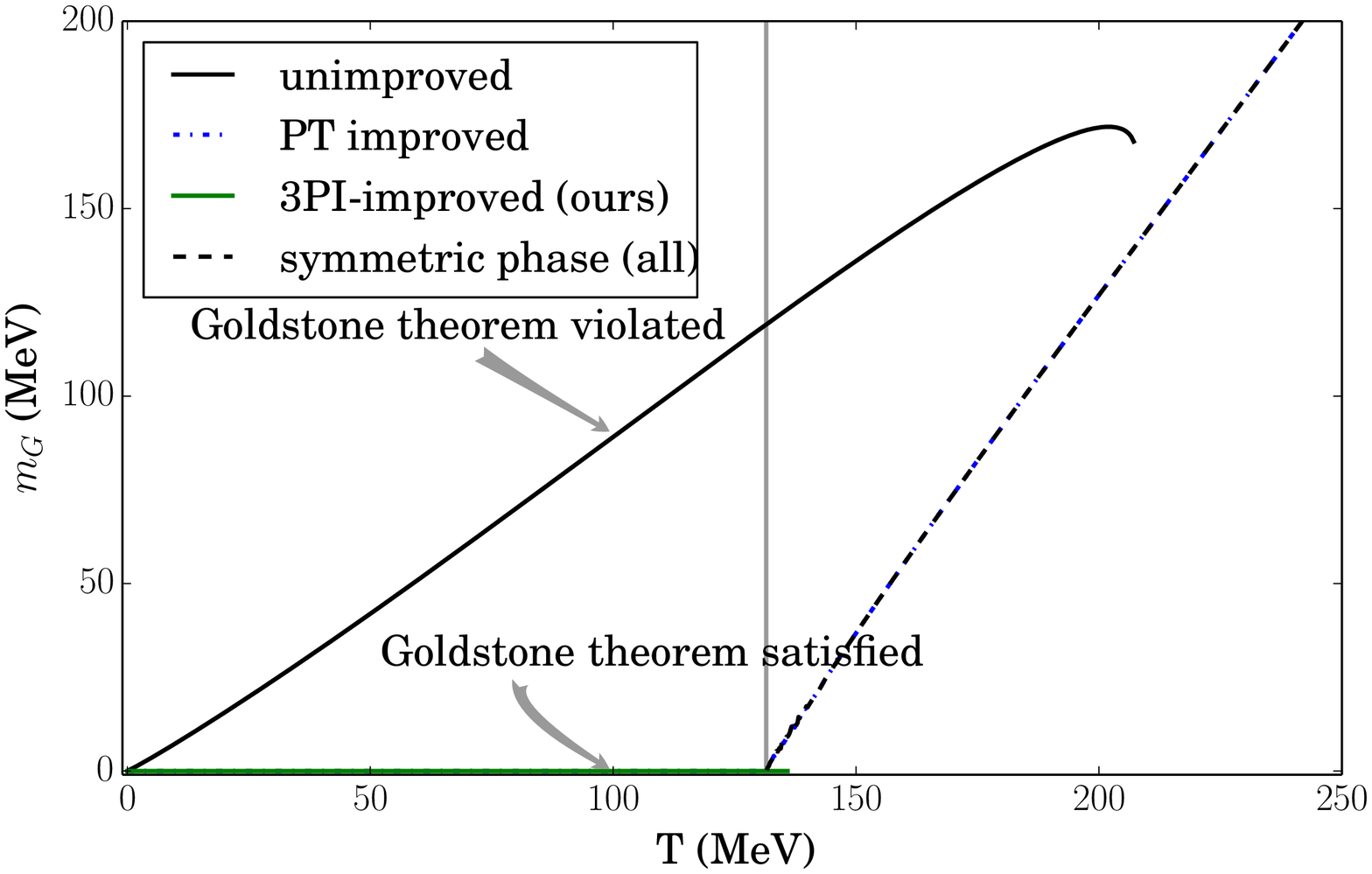}

\protect\caption{\label{fig:hartree-fock-mg-comparison}The Goldstone mass $m_{G}$
as a function of temperature $T$ computed in the Hartree-Fock approximation
using the unimproved 2PIEA (solid black), the Pilaftsis and Teresi
symmetry improved 2PIEA (dash dotted blue) and our symmetry improved
3PIEA (solid green). In the symmetric phase (dashed black) all methods
agree. The vertical grey line at $T\approx131.5\ \mathrm{MeV}$ corresponds
to the critical temperature which is the same in all methods.}
\end{figure}

We present numerical solutions of equations \eqref{eq:mg2-hartree-gap-eq}-\eqref{eq:nosi-vev-eq}
with $N=4$, $v=93\ \mathrm{MeV}$ and $\bar{m}_{H}=500\ \mathrm{MeV}$.
These values are chosen to represent the low energy mesonic sector
of QCD, and to enable direct comparison with \citep{Mao2013}. Our
results are also closely comparable with \citep{Petropoulos2004},
though they take $\bar{m}_{H}\approx600\ \mathrm{MeV}$. The solution
is implemented in \noun{Python} as an iterative root finder based
on \emph{scipy.optimize.root} \citep{*[{}] [{http://www.scipy.org/.}] Jones} with an estimated Jacobian
or, if that fails to converge, a direct iteration of the gap equations.
The Bose-Einstein integrals in \eqref{eq:bose-einstein-integral}
can be precomputed to save time. We show the results for the scalar
field $v$, Higgs mass $m_{H}$ and Goldstone mass $m_{G}$ in Figures
\ref{fig:hartree-fock-vev-comparison}, \ref{fig:hartree-fock-mn-comparison},
and \ref{fig:hartree-fock-mg-comparison} respectively.

Figure \ref{fig:hartree-fock-vev-comparison} shows $v\left(T\right)$,
the order parameter of the phase transition. Below the critical temperature
there is a broken phase with $v\neq0$, but the symmetry is restored
when $v=0$ above the critical temperature. Note, however, that the
unimproved and symmetry improved 3PIEA have unphysical metastable
broken phases at $T>T_{\star}$, signalling a first order phase transition.
The symmetry improved 2PIEA correctly predicts the second order nature
of the phase transition. Though unphysical, the symmetry improved
3PIEA behavior is much more reasonable than the unimproved 2PIEA:
the strength of the first order phase transition is reduced and the
metastable phase ceases to exist at a temperature much closer to the
critical temperature than for the unimproved 2PIEA. Figure \ref{fig:hartree-fock-mn-comparison}
shows the Higgs mass $m_{H}\left(T\right)$. The phase transition
behavior above is seen again, and again all three methods agree in
the symmetric phase, giving the usual thermal mass effect. Finally,
Figure \ref{fig:hartree-fock-mg-comparison} shows the Goldstone boson
mass. The unimproved 2PIEA strongly violates the Goldstone theorem,
but both symmetry improvement methods satisfy it as expected. Note
that the Goldstone theorem is even satisfied in the unphysical metastable
phase predicted by the symmetry improved 3PIEA. All three methods
correctly predict $m_{G}=m_{H}$ in the symmetric phase.

\section{Two dimensions and the Coleman-Mermin-Wagner theorem\label{sub:Two-dimensions-and}}

Recall that the Coleman-Mermin-Wagner theorem \citep{coleman1973},
which has been interpreted as a breakdown of the Goldstone theorem
\citep{Shifman2012}, is a general result stating that the spontaneous
breaking of a continuous symmetry is impossible in $d=2$ or $d=1+1$
dimensions. This occurs due to the infrared divergence of the massless
scalar propagator in two dimensions. We show that the symmetry improved
gap equations satisfy this theorem despite the direct imposition of
Goldstone's theorem. Thus symmetry improvement passes another test
that any robust quantum field theoretical method must satisfy. (Note
that symmetry improvement is not \emph{required} to obtain consistency
of $n$PIEA with the Coleman-Mermin-Wagner theorem, but neither does
it ruin it.)

The general statement of the result is that $\int_{x}\Sigma\left(x,0\right)$
diverges whenever massless particles appear in loops in $d=2$, thus,
by \eqref{eq:dyson-eq} and \eqref{eq:w1-WI-1} $v=0$ and a mass
gap is generated. We will show this explicitly using the Hartree-Fock
gap equations \eqref{eq:mg2-hartree-gap-eq}-\eqref{eq:goldstone-theorem-hartree-gap-eq},
where in two dimensions
\begin{equation}
\mathcal{T}_{a}^{\text{vac}}\left(\overline{\text{MS}}\right)=-\frac{1}{4\pi}\ln\left(\frac{m_{a}^{2}}{\mu^{2}}\right).
\end{equation}
(Note that the renormalization can be carried through without difficulty
in two dimensions. Only the $\delta m_{1}^{2}$ counter-term is needed.)
We must show that the gap equations possess no solution for $m_{G}^{2}=0$.
It is clear that if $v\neq0$, $\mathcal{T}_{G}^{\text{vac}}$ diverges
as $m_{G}^{2}\to0$ if we take $\mu$ as a constant, and $\mathcal{T}_{H}^{\text{vac}}$
diverges if we take $\mu^{2}\propto m_{G}^{2}$ as $m_{G}^{2}\to0$.
Either way there is no solution. At finite temperature the Bose-Einstein
integral $\mathcal{T}_{a}^{\text{th}}$ also has an infrared divergence
as $m_{a}\to0$ which does not cancel against the singularity of the
vacuum term. It can be shown that the singularity is due to the Matsubara
zero mode.

For $v=0$ on the other hand, the gap equations reduce to
\begin{equation}
m_{H}^{2}=m_{G}^{2}=m^{2}-\frac{1}{4\pi}\frac{\hbar}{6}\left(N+2\right)\lambda\ln\left(\frac{m_{G}^{2}}{\mu^{2}}\right),
\end{equation}
which always has a positive solution. If $m^{2}>0$ then one can choose
the renormalization point $\mu^{2}=m_{G}^{2}$ so that the tree level
relationship $m_{G}^{2}=m^{2}$ holds. If $m^{2}<0$ a positive mass
is dynamically generated and one requires a renormalization point
$\mu^{2}>m_{G}^{2}\exp\left(\frac{24\pi\left|m^{2}\right|}{\hbar\left(N+2\right)\lambda}\right)$
non-perturbatively large in the ratio $\lambda/\left|m^{2}\right|$,
reflecting the fact that perturbation theory is bound to fail in this
case.

\section{Optical theorem and dispersion relations\label{sec:Optical-theorem-and}}

\begin{figure}
\includegraphics[width=0.5\columnwidth]{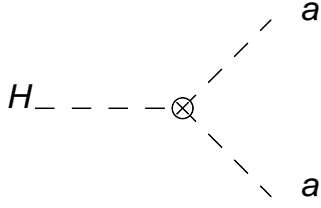}\protect\caption{\label{fig:tree-level-higgs-decay}Tree level of decay of the Higgs
($H$) to two Goldstone bosons $a=1,\cdots,N-1$.}
\end{figure}

In this section we examine the analytic structure of propagators and
self-energies in the symmetry improved 3PI formalism. A physical quantity
of particular interest is the decay width $\Gamma_{H}$ of the Higgs,
which is dominated by decays to two Goldstones. $\Gamma_{H}$ is given
by the optical theorem in terms of the imaginary part of the self-energy
evaluated on-shell (see, e.g. \citep{Peskin1995} Chapter 7):
\begin{equation}
-m_{H}\Gamma_{H}=\mathrm{Im}\Sigma_{H}\left(m_{H}\right).
\end{equation}
(This is valid so long as $\Gamma_{H}\ll m_{H}$, otherwise the full
energy dependence of $\Sigma_{H}\left(p\right)$ must be taken into
account.) The standard one loop perturbative result gives
\begin{equation}
\Gamma_{H}=\frac{N-1}{2}\frac{\hbar}{16\pi m_{H}}\left(\frac{\lambda v}{3}\right)^{2},
\end{equation}
which comes entirely from the Goldstone loop sunset graph. Each part
of this expression has a simple interpretation in relation to the
tree level decay graph (Figure \ref{fig:tree-level-higgs-decay}).
The $N-1$ is due to the sum over final state Goldstone flavours,
the factor of $1/2$ is due to the Bose statistics of the two particles
in the final state, the $\hbar/16\pi m_{H}$ is due to the final state
phase space integration and the $\left(\lambda v/3\right)^{2}$ is
the absolute square of the invariant decay amplitude.

The Hartree-Fock approximation fails to reproduce this result regardless
of the use or not of symmetry improvement. This is because there is
no self-energy apart from a mass correction. Thus the Hartree-Fock
approximation always predicts that the Higgs is stable. Attempts to
repair the Hartree-Fock approximation through the use of an external
propagator lead to a non-zero but still incorrect result. This is
because an unphysical value of $m_{G}$ still appears in loops. 
Satisfactory results are obtained within the symmetry improved 2PI
formalism for both on- and off-shell Higgs \citep{Pilaftsis2013}.
Here we show that the symmetry improved 3PIEA can not yield a satisfactory
value for $\Gamma_{H}$ at the two loop level.

From \eqref{eq:Higgs-eom-finite}
\begin{align}
\mathrm{Im}\Sigma_{H}\left(p\right) & =\frac{\hbar}{6}\left(N+1\right)\lambda\mathrm{Im}\mathcal{T}_{G}^{\text{fin}}+\frac{\hbar}{6}\lambda\mathrm{Im}\mathcal{T}_{H}^{\text{fin}}\nonumber \\
 & -\hbar\left(\frac{\lambda v}{3}\right)^{2}\mathrm{Im}\left[i\mathcal{I}_{HG}^{\text{fin}}\left(p\right)\right]\nonumber \\
 & =\frac{\hbar}{6}\left(N+1\right)\lambda\mathrm{Im}\mathcal{T}_{G}+\frac{\hbar}{6}\lambda\mathrm{Im}\mathcal{T}_{H}\nonumber \\
 & -\hbar\left(\frac{\lambda v}{3}\right)^{2}\mathrm{Im}\left[i\mathcal{I}_{HG}\left(p\right)\right],
\end{align}
which can be written in terms of the un-subtracted $\mathcal{T}_{G/H}$
and $\mathcal{I}_{HG}$ because all of the subtractions are manifestly
real. Now we show that $\mathrm{Im}\mathcal{T}_{G/H}=0$. To do this
we introduce the K\"all\'en-Lehmann spectral representation of the propagators
\citep{Weinberg1995}
\begin{equation}
\Delta_{G/H}\left(q\right)=\int_{0}^{\infty}\mathrm{d}s\frac{\rho_{G/H}\left(s\right)}{q^{2}-s+i\epsilon},
\end{equation}
where the spectral densities $\rho_{G/H}\left(s\right)$ are real and positive
for $s\geq0$ and obey the sum rule
\begin{equation}
\int_{0}^{\infty}\mathrm{d}s\rho_{G/H}\left(s\right)=\left(ZZ_{\Delta}\right)^{-1}=1,
\end{equation}
where the last equality holds at two loop order (we have adapted the
standard formula to our renormalization scheme).

(Note that this standard
theory actually conflicts with the asymptotic $p^{2}\left(\ln p^{2}\right)^{c_{1}}$
form assumed for the self-energy when the $\Phi_{5}$ graph is included,
so that our argument must be refined at the three loop level. The
essential problem is that the self-consistent $n$PI propagator is
not resumming large logarithms. However, it seems unlikely that a
refinement of the argument to account for this fact will change the
qualitative conclusions of this section since, as will be shown shortly,
the predicted $\Gamma_{H}$ is wrong by group theory factors in addition
to the $\mathcal{O}\left(1\right)$ factors which could be compensated
by a modification of $\rho_{G/H}$.)

Then
\begin{equation}
\mathrm{Im}\int_{q}i\int_{0}^{\infty}\mathrm{d}\mu^{2}\frac{\rho_{G/H}\left(\mu^{2}\right)}{q^{2}-\mu^{2}+i\epsilon}=\mathrm{Im}\int_{0}^{\infty}\mathrm{d}\mu^{2}\rho_{G/H}\left(\mu^{2}\right)\mathcal{T}^{\mu}=0.
\end{equation}
This allows us to obtain a dispersion relation relating the real
and imaginary parts of the self-energies
\begin{align}
0 & =\mathrm{Im}\int_{q}i\frac{1}{q^{2}-m_{G/H}^{2}-\Sigma_{G/H}\left(q\right)}\nonumber \\
 & =\int_{q}\frac{q^{2}-m_{G/H}^{2}-\mathrm{Re}\Sigma_{G/H}\left(q\right)}{\left[q^{2}-m_{G/H}^{2}-\mathrm{Re}\Sigma_{G/H}\left(q\right)\right]^{2}+\left[\mathrm{Im}\Sigma_{G/H}\left(q\right)\right]^{2}}.
\end{align}

Finally we have left to compute $\mathrm{Im}\left[i\mathcal{I}_{HG}\left(p\right)\right]$
which can be written
\begin{widetext}
\begin{align}
\mathrm{Im}\left[i\mathcal{I}_{HG}\left(p\right)\right] & =\mathrm{Im}i\int_{q}\int_{0}^{\infty}\mathrm{d}s_{1}\int_{0}^{\infty}\mathrm{d}s_{2}\frac{i\rho_{N}\left(s_{1}\right)}{q^{2}-s_{1}+i\epsilon}\frac{i\rho_{G}\left(s_{2}\right)}{\left(p-q\right)^{2}-s_{2}+i\epsilon}\nonumber \\
 & =\mathrm{Im}i\int_{0}^{\infty}\mathrm{d}s_{1}\int_{0}^{\infty}\mathrm{d}s_{2}\rho_{N}\left(s_{1}\right)\rho_{G}\left(s_{2}\right)\int_{q}\frac{i}{q^{2}-s_{1}+i\epsilon}\frac{i}{\left(p-q\right)^{2}-s_{2}+i\epsilon}\nonumber \\
 & =\frac{1}{16\pi^{2}}\int_{0}^{\infty}\mathrm{d}s_{1}\int_{0}^{\infty}\mathrm{d}s_{2}\rho_{N}\left(s_{1}\right)\rho_{G}\left(s_{2}\right)\mathrm{Im}\int_{0}^{1}\mathrm{d}x\ln\left(\frac{\mu^{2}}{-x\left(1-x\right)p^{2}+xs_{1}+\left(1-x\right)s_{2}-i\epsilon}\right).
\end{align}

\end{widetext}

The imaginary part of the $x$ integral is only nonzero for $\sqrt{s_{1}}+\sqrt{s_{2}}<\sqrt{p^{2}}$.
We denote the region of $s_{1,2}$ integration by $\Omega$. Then the
imaginary part of the $x$ integral can be evaluated straightforwardly,
giving

\begin{align}
\mathrm{Im}\left[i\mathcal{I}_{HG}\left(p\right)\right] & =\frac{1}{16\pi p^{2}}\int_{\Omega}\mathrm{d}s_{1}\mathrm{d}s_{2}\rho_{N}\left(s_{1}\right)\rho_{G}\left(s_{2}\right)\nonumber \\
 & \times\sqrt{p^{2}-\left(\sqrt{s_{1}}+\sqrt{s_{2}}\right)^{2}}\nonumber \\
 & \times\sqrt{p^{2}-\left(\sqrt{s_{1}}-\sqrt{s_{2}}\right)^{2}}.
\end{align}

\begin{figure}
\includegraphics[width=0.5\columnwidth]{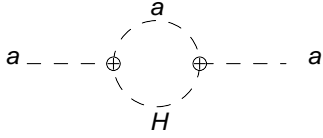}

\protect\caption{\label{fig:si-3pi-higgs-sunset}The self-energy diagram from \eqref{eq:Higgs-eom-finite}
which, due to the inconsistent truncation of the Ward identity,
gives the incorrect absorptive part to the Higgs propagator in the
two loop truncated symmetry improved 3PIEA. $a=1,\cdots,N-1$ labels
Goldstone boson lines and $H$ labels the Higgs boson line.}
\end{figure}

Now, since the each term of the integrand is positive and the square
root is $\leq p^{2}$ we have
\begin{equation}
\mathrm{Im}\left[i\mathcal{I}_{HG}\left(p\right)\right]\leq\frac{1}{16\pi}\int_{\Omega}\mathrm{d}s_{1}\mathrm{d}s_{2}\rho_{N}\left(s_{1}\right)\rho_{G}\left(s_{2}\right)\leq\frac{1}{16\pi},
\end{equation}
using the sum rule for $\rho_{N/G}\left(s\right)$. Finally we have
\begin{equation}
\Gamma_{H}\leq\frac{\hbar}{16\pi m_{H}}\left(\frac{\lambda v}{3}\right)^{2},
\end{equation}
which is smaller than the expected value for all $N>3$. $N=2$ and
$3$ are cases where one could possibly obtain an (accidentally) reasonable
result, depending on the precise form of the spectral functions, but
it is clear that one should not generically expect a correct prediction
of $\Gamma_{H}$ from the symmetry improved 3PIEA at two loops. The
source of the problem is the derivation of the two loop truncation
where we dropped the vertex correction term in \eqref{eq:wi-for-the-self-energies},
resulting in a truncation of the true Ward identity \eqref{eq:w2-WI-1}
that keeps the one loop graphs in $\Sigma_{G}$ but not in $\bar{V}$.
The diagram contributing to $\Gamma_{H}$ above is thus the Goldstone
self-energy shown in Figure \ref{fig:si-3pi-higgs-sunset} which has
the incorrect kinematics and lacks the required group theory $\left(N-1\right)$
and Bose symmetry $\left(1/2\right)$ factors as well. In fact, a
perturbative evaluation of Figure \ref{fig:si-3pi-higgs-sunset} gives
$\Gamma_{H}=0$ due to the threshold at $p^{2}=m_{H}^{2}$! What we
have shown is that no matter the form of the exact spectral functions,
there cannot be a non-perturbative enhancement of this graph large
enough to give the correct $\Gamma_{H}$ for $N>3$. The neglected
vertex corrections give a leading $\mathcal{O}\left(\hbar\right)$
contribution to $\Gamma_{H}$ which must be included.

\begin{figure}
\includegraphics[width=1\columnwidth]{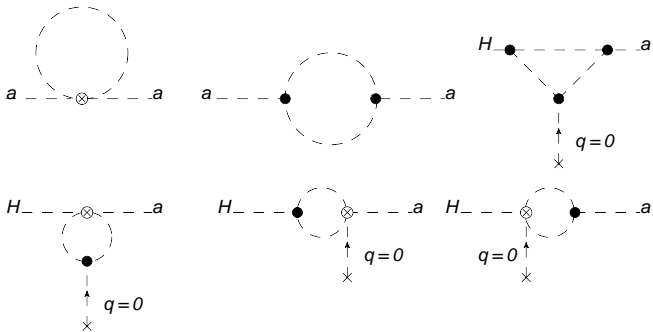}

\protect\caption{\label{fig:one-loop-Higgs-self-energy}One loop contribution to the
Higgs self-energy. The tadpole and sunset graphs are from the Higgs
self-energy $\Sigma_{G}$, while the four remaining terms come from
vertex corrections via the Ward identity \eqref{eq:w2-WI-1}. The
momentum incoming from the lower Goldstone leg is zero, and the crossed
vertex represents a factor of $v$.}
\end{figure}

Now we consider the three loop truncation of the symmetry improved
3PIEA. Since one loop vertex corrections appear at this order we
expect that $\Gamma_{H}$ should be correct at least to $\mathcal{O}\left(\hbar\right)$.
Since the previous result was incorrect by group theory factors already
at $\mathcal{O}\left(\hbar\right)$ our task simplifies to seeking
only the $\mathcal{O}\left(\hbar\right)$ decay width, and so we make
use of only the one loop terms in the Higgs equation of motion, which
are displayed in Figure \ref{fig:one-loop-Higgs-self-energy}. Furthermore,
by iterating the equations of motion we may replace all propagators
and vertices by their perturbative values to $\mathcal{O}\left(\hbar\right)$.
This will leave out contributions of higher order decay processes
such as $H\to GGGG$. We leave the numerical task of computing the
exact decay width predicted by the symmetry improved 3PIEA to future
work.

The contributions of the various terms in Figure \ref{fig:one-loop-Higgs-self-energy}
to the imaginary part of $\Sigma_{H}$ can be determined using Cutkosky
cutting rules \citep{Peskin1995}. In particular, the Hartree-Fock
diagram and the first bubble vertex correction diagram (left diagram,
bottom row Figure \ref{fig:one-loop-Higgs-self-energy}) have no cuts
such that all cut lines can be put on shell. Also, cuts through intermediate
states with both Goldstone and Higgs lines contribute to the process
$H\to HG$, which vanishes due to the zero phase space at threshold.
This means we can drop the sunset diagram and the last bubble vertex
correction (right diagram, bottom row Figure \ref{fig:one-loop-Higgs-self-energy}).
Similarly cuts through two intermediate Higgs lines can be dropped
since $H\to HH$ is impossible on shell. This mean we can drop the
contributions to the triangle and remaining bubble diagram where the
leftmost vertex is $V_{N}$ rather than $\bar{V}$. The contributions
we are interested in can now be displayed explicitly:

\begin{align}
-\Sigma_{H} & \supset\bar{V}v\nonumber \\
 & \supset v\left[i\hbar\left(-\frac{\lambda v}{3}\right)^{3}\int_{\ell}\frac{1}{\left(\ell-p\right)^{2}-m_{G}^{2}+i\epsilon}\right.\nonumber \\
 & \times\frac{1}{\ell^{2}-m_{G}^{2}+i\epsilon}\frac{1}{\ell^{2}-m_{H}^{2}+i\epsilon}\nonumber \\
 & +\frac{i\hbar\lambda}{6}\left(N+1\right)\left(-\frac{\lambda v}{3}\right)\nonumber \\
 & \times\left.\int_{\ell}\frac{1}{\left(\ell-p\right)^{2}-m_{G}^{2}+i\epsilon}\frac{1}{\ell^{2}-m_{G}^{2}+i\epsilon}\right],
\end{align}
where the first and second term are the triangle and bubble graphs
respectively. We now cut the Goldstone lines by replacing each cut
propagator $\left(p^{2}-m_{G}^{2}+i\epsilon\right)^{-1}\to-2\pi i\delta\left(p^{2}-m_{G}^{2}\right)$
to give $-2i\mathrm{Im}\Sigma_{H}$ (because the cutting rules give
the \emph{discontinuity} of the diagram, which is $2i$ times the
imaginary part), yielding

\begin{align}
-2i\mathrm{Im}\Sigma_{H} & \supset-i\hbar v\left[\left(-\frac{\lambda v}{3}\right)^{3}\frac{1}{-m_{H}^{2}}+\frac{\lambda}{6}\left(N+1\right)\left(-\frac{\lambda v}{3}\right)\right]\nonumber \\
 & \times\int_{\ell}2\pi\delta\left(\left(\ell-p\right)^{2}\right)2\pi\delta\left(\ell^{2}\right)\nonumber \\
 & =i\hbar\frac{\lambda^{2}v^{2}}{3^{2}2}\left(N-1\right)\int_{\ell}2\pi\delta\left(\left(\ell-p\right)^{2}\right)2\pi\delta\left(\ell^{2}\right),
\end{align}

The integral can be evaluated by elementary techniques, giving

\begin{align}
\int_{\ell}2\pi\delta\left(\left(\ell-p\right)^{2}\right)2\pi\delta\left(\ell^{2}\right) & =\frac{1}{4\pi^{2}}\int\mathrm{d}^{4}\ell\nonumber \\
 & \times\delta\left(\ell^{2}-2\ell\cdot p+p^{2}\right)\delta\left(\ell^{2}\right)\nonumber \\
 & =\frac{1}{4\pi^{2}}\int\mathrm{d}\ell_{0}\mathrm{d}l4\pi l^{2}\nonumber \\
 & \times\delta\left(-2\ell_{0}m_{H}+m_{H}^{2}\right)\delta\left(\ell_{0}^{2}-l^{2}\right)\nonumber \\
 & =\frac{1}{\pi}\int\mathrm{d}ll^{2}\frac{1}{2m_{H}}\frac{\delta\left(\frac{m_{H}}{2}-l\right)}{2\frac{m_{H}}{2}}\nonumber \\
 & =\frac{1}{8\pi},
\end{align}
and finally

\begin{align}
-\mathrm{Im}\Sigma_{H}\left(m_{H}\right) & =\frac{N-1}{2}\frac{\hbar}{16\pi}\left(\frac{\lambda v}{3}\right)^{2}+\mathcal{O}\left(\hbar^{2}\right).
\end{align}
This exactly matches the expected $\Gamma_{H}$, including group theory
and Bose symmetry factors. The full non-perturbative solution will
give corrections to this accounting for loop corrections as well as
cascade decay processes $H\to GG\to\left(GG\right)^{2}\to\cdots$.
We leave the evaluation of this to future work, however, we have shown
that the one loop vertex corrections are required to get the correct
$\Gamma_{H}$ at leading order.

\section{Discussion\label{sec:Discussion}}

The symmetry improvement formalism of Pilaftsis and Teresi is able
to enforce the preservation of global symmetries in two particle irreducible
effective actions, allowing among other things the accurate description
of phase transitions in strongly coupled theories using numerical
methods that are relatively cheap compared to lattice methods.
As an example of this, during the preparation of this manuscript it was shown
that the symmetry improved 2PIEA solves problems with infrared divergences
of the standard model effective potential due to massless Goldstone bosons \citep{Apo},
though that study was carried out without the gauge sector. It also shows that
the symmetry improved 2PIEA performs better than an \emph{ad hoc} resummation
scheme proposed in the prior literature. This is heartening, though not wholly
surprising due to the inherent self-consistency of $n$PIEA, a topic we plan to
discuss in a forthcoming publication.

However, the development of a first principles non-perturbative kinetic
theory for the gauge theories of real physical interest requires the use
of $n$-particle irreducible effective actions with $n\geq3$. We have taken
a step in this direction by extending the symmetry improvement formalism
to the 3PIEA for a scalar field theory with spontaneous breaking of
a global $\mathrm{O}\left(N\right)$ symmetry. We found that an extra
Ward identity involving the vertex function must be imposed. Since
the constraints are singular this required a careful consideration
of the variational procedure, namely one must be careful to impose
constraints in a way that satisfies d'Alembert's principle. Once this
is done the theory can be renormalized in a more or less standard
way, though the counter-terms differ in value from the unimproved
case. We derived finite equations of motion and counter-terms for
the Hartree-Fock truncation, two loop truncation, and three loop truncation
of the effective action.

We found several important qualitative results. First, symmetry improvement
breaks the equivalence hierarchy of $n$PIEA. Second, the numerical
solution of the Hartree-Fock truncation gave mixed results: Goldstone's
theorem was satisfied, but the order of the phase transition was incorrectly
predicted to be weakly first order (though there was still a large
quantitative improvement over the unimproved 2PIEA case). Third, the
two loop truncation incorrectly predicts the Higgs decay width as
a consequence of the optical theorem, though the three loop truncation
gives the correct value, at least to $\mathcal{O}\left(\hbar\right)$.
These results could be considered strong circumstantial evidence that
one should not apply symmetry improvement to $n$PIEA at a truncation
to less than $n$ loops. One could test this conjecture further by,
for example, computing the symmetry improved 4PIEA. We predict that
unsatisfactory results of some kind will be found for any truncation
of this to $<4$ loops.

Our renormalization of the theory at two and three loops was performed
in vacuum. The only finite temperature computation performed here
was for the Hartree-Fock approximation. The extension of the two or
three loop truncations to finite temperature, or an extension to non-equilibrium
situations, will require a much heavier numerical effort than what
we have attempted. It would also be interesting to compare the self-consistent
Higgs decay rate in the symmetry improved 2PI and 3PI formalisms.
We leave these investigations to future work. Similarly, we presented
analytical results for the renormalization of the three loop truncation
only in $1+2$ dimensions, since the renormalization was not analytically
tractable in $1+3$ dimensions. This is also left to future work.
The general renormalization theory presented here, based on counter-terms,
is difficult to use in practice. It will be interesting to see if
symmetry improvement could work along with the counter-term-free functional
renormalization group approach \citep{Carrington2014}. Such an approach
may not be easier to set up in the first place, but once developed
would likely be easier to extend to higher loop order and $n$ than
the current method. Of course it will be interesting to see if this
work can be extended to gauge symmetries and, eventually, the standard
model of particle physics. If successful, such an effort could serve
to open a new window to the non-perturbative physics of these theories
in high temperature, high density and strong coupling regimes.
\begin{acknowledgments}
We thank Daniele Teresi for clarifying comments concerning \citep{Pilaftsis2013}
and Peter Drummond for useful discussions relating to asymptotic series
and summability.
\end{acknowledgments}

\appendix

\section{The d'Alembert formalism\label{sub:The-d'Alembert-formalism}}

The assumption that $\delta f/\delta\mathcal{W}_{2}$ can be consistently
taken to be constant requires explanation. Constrained Lagrangian problems
are generally under-specified unless one invokes some principle like
d'Alembert's principle (that the constraint forces are ``ideal'',
i.e. they do no work on the system) to specify the constraint forces.
Note that it while it is usually stated that enforcing constraints
through Lagrange multipliers is equivalent to applying d'Alembert's
principle, this is no longer automatically the case if the constraints
involve a singular limit as happens in the field theory case. This
leads to a real ambiguity in the procedure which requires the analyst
to input physical information to resolve it. In the case of mechanical
systems the analyst is expected to be able to furnish the correct
form of the constraints by inspection of the system. However, the
interpretation of ``work'' and ``constraint force'' in the field
theory case is subtle and the appropriate generalization is not
obvious. Here we argue, by way of a simple mechanical analogy, that
the procedure which leads to the maximum simplification of the equations
of motion is the correct field theory analogue of d'Alembert's principle
in mechanics.

d'Alembert's principle is empirically verifiable for a given mechanical
system, but for us it forms part of the \emph{definition} of our approximation
scheme, which we refer to as the ``d'Alembert formalism.'' The result
of Section \ref{sec:Symmetry-improvement} was a set of unambiguous $f$-independent
equations of motion and constraint at some fixed order of the loop
expansion, say $l$-loops. The use of any other limiting procedure
requires the analyst to specify a spacetime function's worth of data
ahead of time, representing the ``work'' that the constraint forces
do. The resulting equations of motion represent a different formulation
of the system and will have a different solution depending on the
choice of ``work'' function.

Imagine that we are competing against another analyst to find the
most accurate solution for a particular system. It is possible that
a competing smart analyst could choose a work function that results
in a more accurate solution than ours, also working at $l$-loops.
However, we could beat the other analyst by working in the d'Alembert
formalism but at higher loop order. We conjecture that the optimum
choice of work function (in the sense of guaranteeing the optimum
accuracy of the resulting solution of the $l$-loop equations) is
merely a clever repackaging of information contained in $>l$-loop
corrections. (We have no proof of this conjecture. Indeed it is hard
to see if any alternative to the d'Alembert formalism is practicable.)
Thus we choose the d'Alembert formalism, which has the virtue of being
a definite procedure requiring little cleverness from the analyst,
at the cost of potentially having a sub-optimal accuracy for a given
loop order.

To illustrate the connection with a mechanics problem consider a classical
particle in 2D constrained to $x^{2}+y^{2}=r^{2}$. The motion is
uniform circular:
\begin{align}
\left(\begin{array}{c}
x\\
y
\end{array}\right) & =r\left(\begin{array}{c}
\cos\left(\omega t+\phi\right)\\
\sin\left(\omega t+\phi\right)
\end{array}\right).
\end{align}
The action is
\begin{align}
S & =\int L\mathrm{d}t-\lambda f\left[w\right],\\
L & =\frac{1}{2}m\left(\dot{x}^{2}+\dot{y}^{2}\right),
\end{align}
where the constraint is $w\left(t\right)=x\left(t\right)^{2}+y\left(t\right)^{2}-r^{2}=0$
and $f\left[w\right]=0$ if $w\left(t\right)=0$. The equations of
motion are
\begin{align}
m\ddot{x}\left(t\right) & =-2\lambda\frac{\delta f}{\delta w\left(t\right)}x\left(t\right),\\
m\ddot{y}\left(t\right) & =-2\lambda\frac{\delta f}{\delta w\left(t\right)}y\left(t\right).
\end{align}

In this mechanics problem we could set $f\left[w\right]=\int w\left(t\right)\mathrm{d}t$
and carry through the problem in the standard way without any complications.
But to mimic the field theory case, where a limiting procedure is
required, we take $f\left[w\right]=\int w\left(t\right)^{2}\mathrm{d}t$.
In this case
\begin{equation}
\frac{\delta f}{\delta w\left(t\right)}=2w\left(t\right)\to0\text{\ as\ }w\to0.
\end{equation}
This requires $\lambda\to\infty$ such that $\lambda\delta f/\delta w$
approaches a finite limit. Importantly, it must approach a $t$ independent
limit, otherwise an unspecified function of time enters the equations
of motion: $m\ddot{x}\left(t\right)=-k\left(t\right)x\left(t\right),$
etc. This limit can be achieved by restricting the class of variations
considered. Let $x\left(t\right)=r\left(t\right)\cos\theta\left(t\right)$
and $y\left(t\right)=r\left(t\right)\sin\theta\left(t\right)$, where
$\delta r\left(t\right)=r\left(t\right)-r$ parametrises deviations
from the constraint surface. Then $w\left(t\right)=2r\delta r\left(t\right)+\mathcal{O}\left(\delta r^{2}\right)$.
We want $\dot{w}\left(t\right)=0$ which is obviously satisfied by
$\delta r\left(t\right)=\delta r$.

We are arguing that we only consider variations of this restricted
form. The variations along the constraint surface (i.e. variations
of $\theta\left(t\right)$) are unrestricted as they should be. Only
variations orthogonal to the constraint surface are restricted. This
is equivalent to d'Alembert's principle. To see this we compute the
second derivative of $r^{2}$ to obtain $\dot{\theta}^{2}-k=\frac{\ddot{r}}{r}.$
When the constraint is enforced $\ddot{r}=0,$ hence $\dot{k}\neq0$
implies $\ddot{\theta}\neq0$: the constraint forces are causing angular
accelerations, doing work on the particle. At constant radius, the
centripetal force only changes if the angular velocity changes.

In the field theory case we have \eqref{eq:si3pi-w2-eom}. For any
given value of $\bar{V}$ and $\Delta_{G}$ only one value of $\Delta_{H}$
satisfies the constraint, given by

\begin{equation}
\Delta_{H}^{-1\star}\left(x,y\right)=\int_{z}\bar{V}\left(x,y,z\right)v+\Delta_{G}^{-1}\left(x,y\right),
\end{equation}
where the $\star$ denotes the constraint solution. This is a holonomic
constraint: in principle we could substitute this into the effective
action directly and not worry about Lagrange multipliers at all (this
is very messy analytically, though it may be numerically feasible).
We suggest that one restrict variations of $\Delta_{H}^{-1}$ to be
of the form $\Delta_{H}^{-1\star}\left(x,y\right)+\delta k$, where
$\delta k$ is a spacetime independent constant. This way we guarantee
\begin{equation}
\frac{\delta f}{\delta\mathcal{W}_{2}\left(x,y\right)}=2\mathcal{W}_{2}\left(x,y\right)=-2\delta k=\text{const},
\end{equation}
and all the desired simplifications go through. Variations of the
other variables are unrestricted. Because the constraint force $B\delta f/\delta\mathcal{W}_{2}$
disappears from the $\Delta_{G}$, $\bar{V}$ and $V_{N}$ equations
of motion the constraint ``does no work'' on these variables, and
the other variables ($v$ and $\Delta_{H}$) are determined solely
by the constraint equations. This seems a fitting field theory analogy
for d'Alembert's principle.

\section{Deriving counter-terms for two loop truncations\label{sec:Deriving-counter-terms-for-2-loops}}
In this section we derive the counter-terms required to renormalize the
3PIEA and equations of motion in the two loop truncation as discussed in Section \ref{sub:Two-loop-truncation}.
Substituting the expressions for bare fields and parameters in terms
of the renormalized fields and parameters according to \eqref{eq:renorm-intro-wavefun-Z}-\eqref{eq:renorm-intro-Z_V} gives the renormalized effective action
\begin{widetext}
\begin{align}
\Gamma^{\left(3\right)} & =\int_{x}\left(-Z_{\Delta}^{-1}\frac{m^{2}+\delta m_{0}^{2}}{2}v^{2}-\frac{\lambda+\delta\lambda_{0}}{4!}v^{4}\right)+\frac{i\hbar}{2}\left(N-1\right)\mathrm{Tr}\ln\left(Z^{-1}Z_{\Delta}^{-1}\Delta_{G}^{-1}\right)+\frac{i\hbar}{2}\mathrm{Tr}\ln\left(Z^{-1}Z_{\Delta}^{-1}\Delta_{H}^{-1}\right)\nonumber \\
 & -\frac{i\hbar}{2}\left(N-1\right)\mathrm{Tr}\left[\left(ZZ_{\Delta}\partial_{\mu}\partial^{\mu}+m^{2}+\delta m_{1}^{2}+Z_{\Delta}\frac{\lambda+\delta\lambda_{1}^{A}}{6}v^{2}\right)\Delta_{G}\right]\nonumber \\
 & -\frac{i\hbar}{2}\mathrm{Tr}\left[\left(ZZ_{\Delta}\partial_{\mu}\partial^{\mu}+m^{2}+\delta m_{1}^{2}+Z_{\Delta}\frac{3\lambda+\delta\lambda_{1}^{A}+2\delta\lambda_{1}^{B}}{6}v^{2}\right)\Delta_{H}\right]+\Gamma_{3}^{\left(3\right)},
\end{align}

\end{widetext}
which agrees with the non-graphical terms of \citep{Pilaftsis2013}
equation (4.4) upon setting $N=2$, dropping an irrelevant constant
$\propto\mathrm{Tr}\ln Z^{-1}$ and noting our different conventions
($m_{\mathrm{here}}^{2}=-m_{\mathrm{PT}}^{2}$ and $\lambda_{\mathrm{here}}=6\lambda_{\mathrm{PT}}$).
The $\delta\lambda$ terms can be derived by substituting $\lambda_{B}\varphi_{Bc}\varphi_{B}^{c}\to Z^{-2}\left(\lambda+\delta\lambda_{1}^{A}\right)Zv^{2}$
and $\lambda_{B}\varphi_{Ba}\varphi_{Bb}\to Z^{-2}\left(\lambda+\delta\lambda_{1}^{B}\right)Zv^{2}\delta_{aN}\delta_{bN}$
into the definition of $\Delta_{0ab}^{-1}$.

The graph functional becomes
\begin{widetext}
\begin{align}
\Gamma_{3}^{\left(3\right)} & =\Phi_{1}-\frac{\hbar^{2}\left(\lambda+\delta\lambda\right)v}{3!}Z_{V}Z_{\Delta}^{3}\int_{xyzw}\Delta_{H}\left(x,y\right)\left[\Delta_{H}\left(x,z\right)\Delta_{H}\left(x,w\right)V_{N}\left(y,z,w\right)\right.\nonumber \\
 & \left.+\left(N-1\right)\Delta_{G}\left(x,z\right)\Delta_{G}\left(x,w\right)\bar{V}\left(y,z,w\right)\right]-\Phi_{2}+\mathcal{O}\left(\hbar^{3}\right),\label{eq:two-loop-graph-functional-with-cts}
\end{align}
with

\begin{align}
\Phi_{1} & =\frac{\hbar^{2}}{24}\left[\left(N+1\right)\lambda+\left(N-1\right)\delta\lambda_{2}^{A}+2\delta\lambda_{2}^{B}\right]Z_{\Delta}^{2}\left(N-1\right)\Delta_{G}\Delta_{G}\nonumber \\
 & +\frac{\hbar^{2}}{24}\left[3\lambda+\delta\lambda_{2}^{A}+2\delta\lambda_{2}^{B}\right]Z_{\Delta}^{2}\Delta_{H}\Delta_{H}+\frac{\hbar^{2}}{24}2\left(\lambda+\delta\lambda_{2}^{A}\right)Z_{\Delta}^{2}\left(N-1\right)\Delta_{G}\Delta_{H}\\
\Phi_{2} & =\frac{\hbar^{2}}{4}\left(N-1\right)Z_{V}^{2}Z_{\Delta}^{3}\bar{V}\bar{V}\Delta_{H}\Delta_{G}\Delta_{G}+\frac{\hbar^{2}}{12}Z_{V}^{2}Z_{\Delta}^{3}V_{N}V_{N}\Delta_{H}\Delta_{H}\Delta_{H}.
\end{align}
\end{widetext}
The $\delta\lambda_{2}^{A/B}$ terms can be found from substituting
$\lambda_{B}\Delta_{Baa}\Delta_{Bbb}\to Z^{-2}\left(\lambda+\delta\lambda_{2}^{A}\right)Z^{2}Z_{\Delta}^{2}\Delta_{aa}\Delta_{bb}$
and $\lambda_{B}\Delta_{Bab}\Delta_{Bba}\to Z^{-2}\left(\lambda+\delta\lambda_{2}^{B}\right)Z^{2}Z_{\Delta}^{2}\Delta_{ab}\Delta_{ba}$
into $\Phi_{1}$. The $\Phi_{1}$ terms correspond to the Hartree-Fock
approximation and agree with the remaining terms of equation (4.4)
of \citep{Pilaftsis2013}. The remaining $\mathcal{O}\left(\hbar^{2}\right)$
terms in $\Gamma_{3}^{\left(3\right)}$ give the sunset diagrams on
replacing $\bar{V}$ and $V_{N}$ by the solution of their equations
of motion at $\mathcal{O}\left(\hbar^{0}\right)$, which give $V_{N}=3\bar{V}=-Z_{V}^{-1}\left(\lambda+\delta\lambda\right)v\times\delta^{\left(4\right)}\left(x-y\right)\delta^{\left(4\right)}\left(x-z\right)$.
We find that $Z_{V}$ cancels on elimination of $\bar{V}$ and $V_{N}$.
It also disappears from the Ward identity once $\bar{V}$ is eliminated
and hence plays no role in the further development.

Going to momentum space the final result is (up to an irrelevant constant)
\begin{widetext}
\begin{align}
\Gamma^{\left(3\right)} & =\int_{x}\left(-Z_{\Delta}^{-1}\frac{m^{2}+\delta m_{0}^{2}}{2}v^{2}-\frac{\lambda+\delta\lambda_{0}}{4!}v^{4}\right)+\frac{i\hbar}{2}\left(N-1\right)\mathrm{Tr}\ln\left(\Delta_{G}^{-1}\right)+\frac{i\hbar}{2}\mathrm{Tr}\ln\left(\Delta_{H}^{-1}\right)\nonumber \\
 & -\frac{i\hbar}{2}\left(N-1\right)\int_{k}\left(-ZZ_{\Delta}k^{2}+m^{2}+\delta m_{1}^{2}+Z_{\Delta}\frac{\lambda+\delta\lambda_{1}^{A}}{6}v^{2}\right)\Delta_{G}\left(k\right)\nonumber \\
 & -\frac{i\hbar}{2}\int_{k}\left(-ZZ_{\Delta}k^{2}+m^{2}+\delta m_{1}^{2}+Z_{\Delta}\frac{3\lambda+\delta\lambda_{1}^{A}+2\delta\lambda_{1}^{B}}{6}v^{2}\right)\Delta_{H}\left(k\right)\nonumber \\
 & +\frac{\hbar^{2}}{24}\left[\left(N+1\right)\lambda+\left(N-1\right)\delta\lambda_{2}^{A}+2\delta\lambda_{2}^{B}\right]Z_{\Delta}^{2}\left(N-1\right)\int_{k}\Delta_{G}\left(k\right)\Delta_{G}\left(k\right)\nonumber \\
 & +\frac{\hbar^{2}}{24}\left[3\lambda+\delta\lambda_{2}^{A}+2\delta\lambda_{2}^{B}\right]Z_{\Delta}^{2}\int_{k}\Delta_{H}\left(k\right)\Delta_{H}\left(k\right)+\frac{\hbar^{2}}{24}2\left(\lambda+\delta\lambda_{2}^{A}\right)Z_{\Delta}^{2}\left(N-1\right)\int_{k}\Delta_{G}\left(k\right)\Delta_{H}\left(k\right)\nonumber \\
 & +\frac{\hbar^{2}}{4}\left[\frac{\left(\lambda+\delta\lambda\right)v}{3}\right]^{2}Z_{\Delta}^{3}\left(N-1\right)\int_{kl}\Delta_{H}\left(k\right)\Delta_{G}\left(l\right)\Delta_{G}\left(k+l\right)\nonumber \\
 & +\frac{\hbar^{2}}{12}\left[\left(\lambda+\delta\lambda\right)v\right]^{2}Z_{\Delta}^{3}\int_{kl}\Delta_{H}\left(k\right)\Delta_{H}\left(l\right)\Delta_{H}\left(k+l\right).
\end{align}
\end{widetext}
From this expression we derive the renormalized equations of motion
\eqref{eq:two-loop-Delta_G-eom-with-cts}-\eqref{eq:two-loop-Goldstone-thm-with-cts}

The divergent integrals $\mathcal{T}_{G/H}$ \eqref{eq:tadpole-integral}
and $\mathcal{I}_{HG}\left(p\right)$ \eqref{eq:Ing-integral} enter
into the equations of motion. $\mathcal{I}_{HG}\left(p\right)$ can
be rendered finite by a single subtraction
\begin{equation}
\mathcal{I}_{HG}\left(p\right)=\mathcal{I}^{\mu}+\mathcal{I}_{HG}^{\text{fin}}\left(p\right),\label{eq:Ifinng}
\end{equation}
where $\mathcal{I}^{\mu}=\int_{q}\left[i\Delta^{\mu}\left(q\right)\right]^{2}$.
Since we wrote the propagators with the physical masses explicit,
it is crucial to also subtract a portion of the finite piece $\mathcal{I}_{HG}^{\text{fin}}\left(m_{G/H}\right)$
so that the pole of the propagator is fixed at the physical mass of
the Goldstone/Higgs propagator respectively. We make this subtraction
separately so as to have a universal $\mathcal{I}^{\mu}$.

The tadpole integrals $\mathcal{T}_{G/H}$ require two subtractions
each since $\int_{q}i\left[\Delta^{\mu}\left(q\right)\right]^{2}\Sigma_{G/H}^{0}\left(q\right)$
is logarithmically divergent. To that end we introduce 
\begin{align}
\Sigma^{\mu}\left(q\right) & =-i\hbar\left[\frac{\left(\lambda+\delta\lambda\right)v}{3}\right]^{2}Z_{\Delta}^{3}\nonumber \\
 & \times\left[\int_{\ell}i\Delta^{\mu}\left(\ell\right)i\Delta^{\mu}\left(q+\ell\right)-\mathcal{I}^{\mu}\right],
\end{align}
which is asymptotically the same as $\Sigma_{G/H}^{0}\left(q\right)$,
so that $\int_{q}i\left[\Delta^{\mu}\left(q\right)\right]^{2}\left[\Sigma_{G/H}^{0}\left(q\right)-\Sigma^{\mu}\left(q\right)\right]$
is finite. For later convenience we write
\begin{equation}
\int_{q}i\left[\Delta^{\mu}\left(q\right)\right]^{2}\Sigma^{\mu}\left(q\right)=\hbar\left[\frac{\left(\lambda+\delta\lambda\right)v}{3}\right]^{2}Z_{\Delta}^{3}c^{\mu}.
\end{equation}
Then
\begin{align}
\mathcal{T}_{G/H} & =\mathcal{T}^{\mu}-i\left(m_{G/H}^{2}-\mu^{2}\right)\mathcal{I}^{\mu}\nonumber \\
 & +\hbar\left[\frac{\left(\lambda+\delta\lambda\right)v}{3}\right]^{2}Z_{\Delta}^{3}c^{\mu}+\mathcal{T}_{G/H}^{\text{fin}},\label{eq:Tfing}
\end{align}
where $\mathcal{T}^{\mu}=\int_{q}i\Delta^{\mu}\left(q\right)$. Note
that $\mathcal{T}^{\mu}$ and $c^{\mu}$ are real and $\mathcal{I}^{\mu}$
is imaginary, so that all of the subtractions can be absorbed into
real counter-terms.

The counter-terms are found by eliminating $m_{G/H}^{2}$ and demanding
that the divergences proportional to different powers of $v^{2}$
and $\mathcal{T}_{G/H}^{\text{fin}}$ separately vanish. Further,
we enforce $\Delta_{G}^{-1}\left(m_{G}\right)=0$ and $\Delta_{H}^{-1}\left(m_{H}\right)=0$
and that the counter-terms are momentum independent. This gives eight
equations for the seven constants $Z,Z_{\Delta},\delta m_{1}^{2},\delta\lambda_{1}^{A},\delta\lambda_{2}^{A},\delta\lambda_{2}^{B}$
and $\delta\lambda$, however one of them is redundant and a solution
exists \citep{supp}.

We find nontrivial field strength renormalizations
\begin{equation}
Z=Z_{\Delta}^{-1}=\left\{ 1+\frac{i\hbar\lambda}{3}\left[\mathcal{I}_{HG}^{\text{fin}}\left(m_{H}\right)-\mathcal{I}_{HG}^{\text{fin}}\left(m_{G}\right)\right]\right\} ^{2},
\end{equation}
and a nonzero 
\begin{equation}
\delta\lambda=-\lambda\pm\lambda\left\{ 1+\frac{i\hbar\lambda}{3}\left[\mathcal{I}_{HG}^{\text{fin}}\left(m_{H}\right)-\mathcal{I}_{HG}^{\text{fin}}\left(m_{G}\right)\right]\right\} ^{3}
\end{equation}
 (the two solutions arise because $\delta\lambda$ only enters the
equations of motion in the quadratic combination $\left(\lambda+\delta\lambda\right)^{2}$).
These counter-terms are normally trivial ($Z=Z_{\Delta}=1$ and $\delta\lambda=0$)
for $\phi^{4}$ theory at two loops. However, due to the modification
of the Higgs equation of motion, we require $Z_{\Delta}\neq1$ in
order to enforce $\Delta_{H}^{-1}\left(m_{H}\right)=0$ and this is
then compensated by $Z$ and $\delta\lambda$ in order to recover
the other renormalization conditions. The other counter-terms can
be obtained for any regulator, but the expressions are bulky and unenlightening
even for dimensional regularization in $d=4-2\epsilon$ dimensions,
so we leave their explicit forms in the supplemental \noun{Mathematica}
notebook.

\section{Auxiliary vertex and renormalization in 3 and 4 dimensions\label{sec:Auxiliary-vertex-and-renorm}}

As described in Section \ref{sub:Three-loop-truncation} the renormalization
of the three loop 3PIEA requires the definition of an auxiliary vertex $V^{\mu}_{abc}$ with
the same asymptotic behaviour as the full self-consistent solution at large
momentum. This auxiliary vertex can be found in terms of a six point kernel
which obeys the integral equation \eqref{eq:K-6pt-kernel-eom} illustrated
in Figure \ref{fig:aux-vertex-soln-and-6pt-kernel}..

\begin{figure}
\includegraphics[width=1\columnwidth]{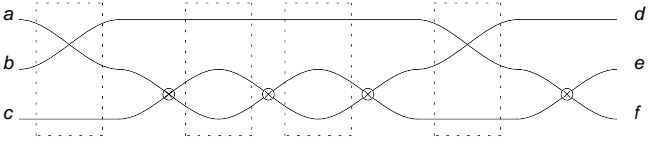}

\protect\caption{\label{fig:K-6pt-contribution-derangement-stab-stab-derangement}A
contribution to the six point kernel $\mathcal{K}_{abcdef}$ resulting
from (left to right) a derangement, two stabilizers and another derangement.
The dashed boxes surround the permutations (to aid visualization only).
Only stabilizers lead to divergent loops.}
\end{figure}

Solving \eqref{eq:K-6pt-kernel-eom} by iteration generates an infinite
number of terms, one of which is illustrated in Figure
\ref{fig:K-6pt-contribution-derangement-stab-stab-derangement}.
Each contribution is in one-to-one correspondence with the
sequence of permutations $\pi_{1}\pi_{2}\cdots\pi_{n}\cdots$ of the propagator lines
(read from left to right in relation to the diagram). Now we divide the
permutations into two classes: ``stabilizers,'' for which $\pi\left(a\right)=a$,
and ``derangements,'' for which $\pi\left(a\right)=b$ or $c$.

\begin{figure}
\includegraphics[width=1\columnwidth]{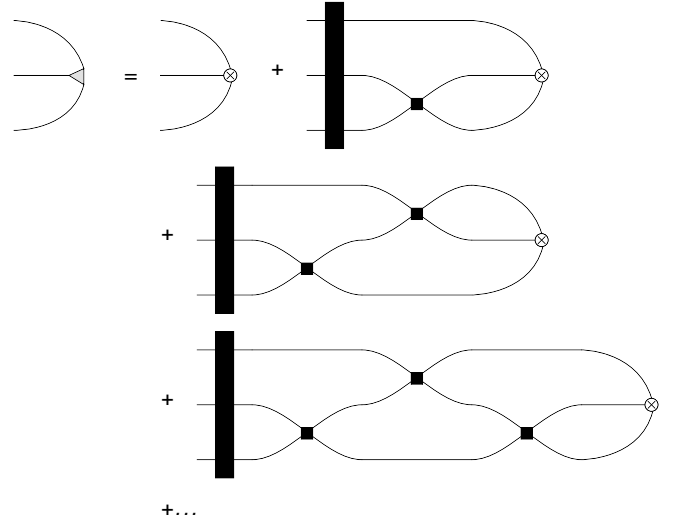}

\protect\caption{\label{fig:aux-vertex-soln}Solution for the auxiliary vertex $V^{\mu}_{abc}$
in terms of the four point kernel which sums all iterated bubble insertions.}
\end{figure}

Any sequence of permutations is of the form of an alternating sequence
of runs of (possibly zero) stabilizers, separated by derangements.
Consider a run of $n$ stabilizers, $\cdots\pi_{a}\left(\pi_{1}\pi_{2}\cdots\pi_{n}\right)\pi_{b}\cdots$,
where $\pi_{a}$ and $\pi_{b}$ are derangements and $\pi_{1}$ through
$\pi_{n}$ are all stabilizers. The case for $n=2$ is shown in Figure
\ref{fig:K-6pt-contribution-derangement-stab-stab-derangement}. Each
stabilizer creates a logarithmically divergent loop on the bottom
two lines $\sim-\lambda\mathcal{I}^{\mu}$. Derangements on the other
hand, if they create loops at all, create loops with $>2$ propagators,
and hence are convergent. Thus all divergences in $\mathcal{K}_{abcdef}$
can be removed by rendering a single primitive divergence finite.
Note that the whole series $\sum_{n=0}^{\infty}\cdots\pi_{a}\left(\prod_{i=1}^{n}\pi_{i}\right)\pi_{b}$,
where again $\pi_{a,b}$ are derangements and $\left\{ \pi_{i}\right\} $
are stabilizers, can be summed because the series is geometric.
The result is that the six point kernel can be determined by an equation
like \eqref{eq:K-6pt-kernel-eom}, except that the sum over all permutations
is replaced by a sum over derangements only, and the bare vertex $W$ is replaced
by a four point kernel $\mathcal{K}_{abcd}^{\left(4\right)}\sim\lambda/\left(1+\lambda\mathcal{I}^{\mu}\right)$.
Denoting this four point kernel by a square vertex we can finally
write the solution for $V^{\mu}_{abc}$ in Figure \ref{fig:aux-vertex-soln}.

This expression for $V^{\mu}_{abc}$ can be dramatically simplified in $3$
or $1+2$ dimensions because $\mathcal{I}^{\mu}$ is finite and the geometric
sum in $\mathcal{K}_{abcd}$ converges. Indeed
$\mathcal{K}_{abcd}^{\left(4\right)}\left(p_{1},p_{2},p_{3},p_{1}+p_{2}-p_{3}\right)\sim\lambda/\left[1+\lambda/\left(p_{1}+p_{2}\right)^{4-d}\right]\to\lambda$
as $p_{1,2,3,4}\to\infty$. Further, every loop integral in Figure \ref{fig:aux-vertex-soln}
likewise converges, and every loop yields a factor of $\sim1/p^{4-d}$.
Thus the dominant behaviour as $p\to\infty$ is just the tree level behaviour
and we can eliminate the auxiliary vertex completely.

However, in $4$ or $1+3$ dimensions $V^{\mu}_{abc}$ apparently cannot be
simplified further. First $\mathcal{K}_{abcd}^{\left(4\right)}$ must
be renormalized, then the bubble appearing in the nontrivial terms
in Figure \ref{fig:aux-vertex-soln} (or the equivalent integral equation)
must be renormalized, then the resulting series must be summed (or
the equivalent integral equation solved), noting that on the basis
of power counting every term is apparently equally important. On this
basis we expect that no compact analytic expression for $V^{\mu}_{abc}$,
or even its asymptotic behavior, exists and that the renormalization
must be accomplished as part of the self-consistent numerical solution
of the full equations of motion.

This style of argument can be quickly generalized to many other theories,
such as gauge theories, where the diagrammatic expansion has a similar
combinatorial structure to scalar $O\left(N\right)$ theory, showing up the well
known problem of the renormalization of $n$PIEA for $n\geq 3$ in four dimensions.
The discussion here certainly does not solve this problem, which remains open,
to our knowledge, though we hope this discussion may be helpful.

\section{Deriving counter-terms for three loop truncations\label{sec:Deriving-Counter-terms-for-3-loops}}

In this section we work in $1+2$ dimensions as discussed in Section
\ref{sub:Three-loop-truncation}. The effective action is as in Appendix
\ref{sec:Deriving-counter-terms-for-2-loops} (before eliminating
$\bar{V}$ and $V_{N}$) except we introduce a new counter-term $\delta\lambda\to\delta\lambda_{C}$
for the second term in \eqref{eq:two-loop-graph-functional-with-cts}
and add the three loop diagrams
\begin{widetext}
\begin{align}
\Phi_{3} & =Z_{V}^{4}Z_{\Delta}^{6}\left[\left(N-1\right)\frac{i\hbar^{3}}{3!}V_{N}\left(\bar{V}\right)^{3}\left(\Delta_{H}\right)^{3}\left(\Delta_{G}\right)^{3}+\frac{i\hbar^{3}}{4!}\left(V_{N}\right)^{4}\left(\Delta_{H}\right)^{6}+\left(N-1\right)\frac{i\hbar^{3}}{8}\left(\bar{V}\right)^{4}\Delta_{H}\Delta_{H}\left(\Delta_{G}\right)^{4}\right],\\
\Phi_{4} & =\frac{i\hbar^{3}\left(\lambda+\delta\lambda\right)}{24}Z_{V}^{2}Z_{\Delta}^{5}\left[2\left(N-1\right)\bar{V}V_{N}\left(\Delta_{H}\right)^{3}\Delta_{G}\Delta_{G}\right.+\left(N^{2}-1\right)\bar{V}\bar{V}\Delta_{H}\left(\Delta_{G}\right)^{4}+3V_{N}V_{N}\left(\Delta_{H}\right)^{5}\nonumber \\
 & \left.+2^{2}\left(N-1\right)\bar{V}\bar{V}\left(\Delta_{G}\right)^{3}\Delta_{H}\Delta_{H}\right],\\
\Phi_{5} & =\frac{i\hbar^{3}\left(\lambda+\delta\lambda\right)^{2}}{144}Z_{\Delta}^{4}\left\{ \left[\left(N-1\right)\Delta_{G}\Delta_{G}+\Delta_{H}\Delta_{H}\right]^{2}+2\left(N-1\right)\left(\Delta_{G}\right)^{4}+2\left(\Delta_{H}\right)^{4}\right\} .
\end{align}

The equations of motion following from $\Gamma^{\left(3\right)}$
are then

\begin{align}
\Delta_{G}^{-1} & =-\left(ZZ_{\Delta}\partial_{\mu}\partial^{\mu}+m^{2}+\delta m_{1}^{2}+Z_{\Delta}\frac{\lambda+\delta\lambda_{1}^{A}}{6}v^{2}\right)\nonumber \\
 & -\frac{\hbar}{6}\left[\left(N+1\right)\lambda+\left(N-1\right)\delta\lambda_{2}^{A}+2\delta\lambda_{2}^{B}\right]Z_{\Delta}^{2}\mathcal{T}_{G}-\frac{\hbar}{6}\left(\lambda+\delta\lambda_{2}^{A}\right)Z_{\Delta}^{2}\mathcal{T}_{H}\nonumber \\
 & -i\hbar Z_{V}^{2}Z_{\Delta}^{3}\left[-2\frac{\left(\lambda+\delta\lambda_{C}\right)Z_{V}^{-1}v}{3}-\bar{V}\right]\Delta_{H}\Delta_{G}\bar{V}\nonumber \\
 & +\hbar^{2}Z_{V}^{4}Z_{\Delta}^{6}\left[V_{N}\left(\bar{V}\right)^{3}\left(\Delta_{H}\right)^{3}\left(\Delta_{G}\right)^{2}+\left(\bar{V}\right)^{4}\Delta_{H}\Delta_{H}\left(\Delta_{G}\right)^{3}\right]\nonumber \\
 & +\frac{\hbar^{2}\left(\lambda+\delta\lambda\right)}{3}Z_{V}^{2}Z_{\Delta}^{5}\left[\bar{V}V_{N}\left(\Delta_{H}\right)^{3}\Delta_{G}+\left(N+1\right)\bar{V}\bar{V}\Delta_{H}\left(\Delta_{G}\right)^{3}+3\bar{V}\bar{V}\left(\Delta_{G}\right)^{2}\Delta_{H}\Delta_{H}\right]\nonumber \\
 & +\frac{\hbar^{2}\left(\lambda+\delta\lambda\right)^{2}}{18}Z_{\Delta}^{4}\left[\left(N+1\right)\left(\Delta_{G}\right)^{3}+\Delta_{H}\Delta_{H}\Delta_{G}\right],\label{eq:three-loop-Delta_G-eom}
\end{align}
for the Goldstone propagator,

\begin{align}
\bar{V} & =-\frac{\left(\lambda+\delta\lambda_{C}\right)v}{3}Z_{V}^{-1}\nonumber \\
 & +i\hbar Z_{V}^{2}Z_{\Delta}^{3}\left[V_{N}\left(\bar{V}\right)^{2}\left(\Delta_{H}\right)^{2}\Delta_{G}+\left(\bar{V}\right)^{3}\Delta_{H}\left(\Delta_{G}\right)^{2}\right]\nonumber \\
 & +\frac{i\hbar\left(\lambda+\delta\lambda\right)}{6}Z_{\Delta}^{2}\left[V_{N}\left(\Delta_{H}\right)^{2}+\left(N+1\right)\bar{V}\left(\Delta_{G}\right)^{2}+4\bar{V}\Delta_{G}\Delta_{H}\right],\label{eq:three-loop-Vbar-eom}
\end{align}
for the Higgs-Goldstone-Goldstone vertex,

\begin{align}
V_{N} & =-\left(\lambda+\delta\lambda_{C}\right)vZ_{V}^{-1}\nonumber \\
 & +i\hbar Z_{V}^{2}Z_{\Delta}^{3}\left[\left(N-1\right)\left(\bar{V}\right)^{3}\left(\Delta_{G}\right)^{3}+\left(V_{N}\right)^{3}\left(\Delta_{H}\right)^{3}\right]\nonumber \\
 & +\frac{i\hbar\left(\lambda+\delta\lambda\right)}{2}Z_{\Delta}^{2}\left[\left(N-1\right)\bar{V}\Delta_{G}\Delta_{G}+3V_{N}\left(\Delta_{H}\right)^{2}\right],\label{eq:three-loop-V_N-eom}
\end{align}

\end{widetext}
for the triple Higgs vertex, and finally

\begin{align}
0 & =\Delta_{G}^{-1}\left(p=0\right)v,\label{eq:WI-1-three-loop}\\
0 & =Z_{V}Z_{\Delta}\bar{V}\left(p,-p,0\right)v+\Delta_{G}^{-1}\left(p\right)-\Delta_{H}^{-1}\left(p\right),\label{eq:WI2-three-loop}
\end{align}
for the Ward identities.

Note that the only divergent integrals in these equations are the
linearly divergent tadpole integrals $\mathcal{T}_{G/H}$ and the
logarithmically divergent \noun{BBALL} integrals (last line of \eqref{eq:three-loop-Delta_G-eom}).
By power counting with reference to Figure \ref{fig:two-loop-self-energy-graphs}
one finds that the third, fourth, and fifth lines of \eqref{eq:three-loop-Delta_G-eom}
produce finite self-energy contributions with leading asymptotics
$\sim p^{-1}$, $p^{-4}$, and $p^{-2}$ respectively. We can separate
finite and divergent parts of $\Delta_{G}^{-1}$ as
\begin{align}
\Delta_{G}^{-1} & =-\left(\partial_{\mu}\partial^{\mu}+m^{2}+\frac{\lambda}{6}v^{2}\right)\nonumber \\
 & -\left[\Sigma_{G}^{0}\left(p\right)-\Sigma_{G}^{0}\left(m_{G}\right)\right]-\Sigma_{G}^{\infty}\left(p\right),
\end{align}
where
\begin{widetext}
\begin{align}
-\Sigma_{G}^{0}\left(p\right) & =-\frac{\hbar}{6}\left(N+1\right)\lambda\left(\mathcal{T}_{G}-\mathcal{T}^{\mu}\right)-\frac{\hbar}{6}\lambda\left(\mathcal{T}_{H}-\mathcal{T}^{\mu}\right)\nonumber \\
 & -i\hbar\left[-2\frac{\left(\lambda+\delta\lambda_{C}\right)Z_{V}^{-1}v}{3}-\bar{V}\right]\Delta_{H}\Delta_{G}\bar{V}+\hbar^{2}\left[V_{N}\left(\bar{V}\right)^{3}\left(\Delta_{H}\right)^{3}\left(\Delta_{G}\right)^{2}+\left(\bar{V}\right)^{4}\Delta_{H}\Delta_{H}\left(\Delta_{G}\right)^{3}\right]\nonumber \\
 & +\frac{\hbar^{2}\left(\lambda+\delta\lambda\right)Z_{\Delta}^{2}}{3}\left[\bar{V}V_{N}\left(\Delta_{H}\right)^{3}\Delta_{G}+\left(N+1\right)\bar{V}\bar{V}\Delta_{H}\left(\Delta_{G}\right)^{3}+3\bar{V}\bar{V}\left(\Delta_{G}\right)^{2}\Delta_{H}\Delta_{H}\right]\nonumber \\
 & +\frac{\hbar^{2}\left(\lambda+\delta\lambda\right)^{2}Z_{\Delta}^{4}}{18}\left[\left(N+1\right)\left(\Delta_{G}\right)^{3}+\Delta_{H}\Delta_{H}\Delta_{G}-\left(N+2\right)\mathcal{B}^{\mu}\right],
\end{align}

and

\begin{align}
-\Sigma_{G}^{\infty}\left(p\right) & =-\Sigma_{G}^{0}\left(m_{G}\right)-\left(\left(ZZ_{\Delta}-1\right)\partial_{\mu}\partial^{\mu}+\delta m_{1}^{2}+\frac{\delta\lambda_{1}^{A}}{6}v^{2}+\left(Z_{\Delta}-1\right)\frac{\lambda+\delta\lambda_{1}^{A}}{6}v^{2}\right)\nonumber \\
 & -\frac{\hbar}{6}\left(N+1\right)\lambda\mathcal{T}^{\mu}-\frac{\hbar}{6}\left[\left(N-1\right)\delta\lambda_{2}^{A}+2\delta\lambda_{2}^{B}\right]\mathcal{T}_{G}-\frac{\hbar}{6}\left[\left(N+1\right)\lambda+\left(N-1\right)\delta\lambda_{2}^{A}+2\delta\lambda_{2}^{B}\right]\left(Z_{\Delta}^{2}-1\right)\mathcal{T}_{G}\nonumber \\
 & -\frac{\hbar}{6}\lambda\mathcal{T}^{\mu}-\frac{\hbar}{6}\delta\lambda_{2}^{A}\mathcal{T}_{H}-\frac{\hbar}{6}\left(\lambda+\delta\lambda_{2}^{A}\right)\left(Z_{\Delta}^{2}-1\right)\mathcal{T}_{H}\nonumber \\
 & -i\hbar\left(Z_{V}^{2}Z_{\Delta}^{3}-1\right)\left[-2\frac{\left(\lambda+\delta\lambda_{C}\right)Z_{V}^{-1}v}{3}-\bar{V}\right]\Delta_{H}\Delta_{G}\bar{V}\nonumber \\
 & +\hbar^{2}\left(Z_{V}^{4}Z_{\Delta}^{6}-1\right)\left[V_{N}\left(\bar{V}\right)^{3}\left(\Delta_{H}\right)^{3}\left(\Delta_{G}\right)^{2}+\left(\bar{V}\right)^{4}\Delta_{H}\Delta_{H}\left(\Delta_{G}\right)^{3}\right]\nonumber \\
 & +\frac{\hbar^{2}\left(\lambda+\delta\lambda\right)Z_{\Delta}^{2}}{3}\left(Z_{V}^{2}Z_{\Delta}^{3}-1\right)\left[\bar{V}V_{N}\left(\Delta_{H}\right)^{3}\Delta_{G}+\left(N+1\right)\bar{V}\bar{V}\Delta_{H}\left(\Delta_{G}\right)^{3}+3\bar{V}\bar{V}\left(\Delta_{G}\right)^{2}\Delta_{H}\Delta_{H}\right]\nonumber \\
 & +\left(N+2\right)\frac{\hbar^{2}\left(\lambda+\delta\lambda\right)^{2}Z_{\Delta}^{4}}{18}\mathcal{B}^{\mu},
\end{align}

\end{widetext}
are the finite and divergent parts respectively and we introduced
the \noun{bball} integral $\mathcal{B}^{\mu}=\int_{qp}\Delta^{\mu}\left(q\right)\Delta^{\mu}\left(p\right)\Delta^{\mu}\left(p+q\right)$.
In this split we have already assumed that $\left(\lambda+\delta\lambda_{C}\right)Z_{V}^{-1}$
and $\left(\lambda+\delta\lambda\right)Z_{\Delta}^{2}$ are finite,
which will turn out to be the case. Renormalization requires $\Sigma_{G}^{\infty}\left(p\right)=0$.
Note the explicit subtraction of $\Sigma_{G}^{0}\left(m_{G}\right)$
in order to fulfill the mass shell condition. Doing the same now for
$\Delta_{H}^{-1}$ we find the pole condition
\begin{equation}
0=Z_{V}Z_{\Delta}\bar{V}\left(m_{H},-m_{H},0\right)v+m_{H}^{2}-m_{G}^{2}-\Sigma_{G}^{0}\left(m_{H}\right),
\end{equation}
which requires
\begin{equation}
Z_{V}Z_{\Delta}=\frac{m_{G}^{2}+\Sigma_{G}^{0}\left(m_{H}\right)-m_{H}^{2}}{\bar{V}\left(m_{H},-m_{H},0\right)v}\equiv\kappa,
\end{equation}
which is finite. We take for our other renormalization conditions
the separate vanishing of kinematically independent divergences, implying
\begin{align}
ZZ_{\Delta} & =1,\\
\delta m_{1}^{2} & =-\Sigma_{G}^{0}\left(m_{G}\right)-\frac{\hbar}{6}\left(N+2\right)\lambda\mathcal{T}^{\mu}+\left(N+2\right)\frac{\hbar^{2}\lambda^{2}}{18}\mathcal{B}^{\mu},\\
\delta\lambda_{1}^{A} & =-\frac{\left(Z_{\Delta}-1\right)\lambda}{Z_{\Delta}}\\
Z_{V}^{2}Z_{\Delta}^{3} & =1,\\
0 & =\left(N-1\right)\delta\lambda_{2}^{A}+2\delta\lambda_{2}^{B}\nonumber \\
 & +\left[\left(N+1\right)\lambda+\left(N-1\right)\delta\lambda_{2}^{A}+2\delta\lambda_{2}^{B}\right]\left(Z_{\Delta}^{2}-1\right),\\
0 & =\delta\lambda_{2}^{A}+\left(\lambda+\delta\lambda_{2}^{A}\right)\left(Z_{\Delta}^{2}-1\right).
\end{align}
We also choose the conditions
\begin{align}
\left(\lambda+\delta\lambda\right)Z_{\Delta}^{2} & =\lambda,\\
\left(\lambda+\delta\lambda_{C}\right)Z_{V}^{-1} & =\lambda,
\end{align}
to recover the tree level asymptotics for $\bar{V}$ and $V_{N}$.
These conditions give a closed system of nine equations for the nine
quantities $Z$, $Z_{\Delta}$, $Z_{V}$, $\delta m_{1}^{2}$, $\delta\lambda_{1}^{A}$,
$\delta\lambda_{2}^{A/B}$, $\delta\lambda$, and $\delta\lambda_{C}$.

These conditions determine
\begin{align}
\delta\lambda_{2}^{A} & =\delta\lambda_{2}^{B}=-\frac{\lambda\left(Z_{\Delta}^{2}-1\right)}{Z_{\Delta}^{2}}=\left(\kappa^{2}-1\right)\lambda,\\
Z_{V} & =\kappa^{3},\\
Z_{\Delta} & =\kappa^{-2},\\
Z & =\kappa^{2},\\
\delta\lambda & =\left(\kappa^{4}-1\right)\lambda,\\
\delta\lambda_{C} & =\left(\kappa^{3}-1\right)\lambda.
\end{align}

Note that if $\kappa=1$ all of the counter-terms except $\delta m_{1}^{2}$
vanish. This is a manifestation of the super-renormalizability of
$\phi^{4}$ theory in $1+2$ dimensions. The non-zero, indeed finite,
values of all of the other counter-terms are not required to UV-renormalize
the theory, but only to maintain the pole condition for the Higgs
propagator despite the vertex Ward identity.

%\bibliographystyle{apsrev4-1}
%\bibliography{../refs}


\end{document}